\documentclass[apj]{emulateapj}

\usepackage[]{hyperref}
\usepackage[english]{babel} 
\usepackage{longtable}
\usepackage{alltt} 
\usepackage{booktabs} 
\usepackage{multirow} 
\usepackage{overpic} 
\usepackage{xcolor}
\usepackage{seqsplit} 
\usepackage[caption=false]{subfig} 
\usepackage{units} 
\usepackage{pict2e} 
\usepackage{soul} 

\newcommand{\Sersic}{S\'{e}rsic}
\newcommand{\corser}{core-S\'{e}rsic}
\newcommand{\chisq}{$\chi^{2}$}
\newcommand{\GALFIT}{\textsc{Galfit}}
\newcommand{\CORSAIR}{\textsc{Galfit-Corsair}}
\newcommand{\MBH}{$M_{\bullet}$}
\newcommand{\Msph}{$M_{\rm sph,*}$}
\newcommand{\Msun}{$M_{\odot}$}
\newcommand{\magsb}{mag arcsec$^{-2}$}

\shorttitle{The quest for the largest depleted galaxy core}
\shortauthors{P.\ Bonfini \& A.\ W.\ Graham} 

\begin{document}

\title{
 The quest for the largest depleted galaxy core: supermassive black hole binaries
 and stalled in-falling satellites
}

\author{Paolo Bonfini$^{\star}$}
\affil{Centre for Astrophysics and Supercomputing, Swinburne University of
 Technology, Victoria 3122, Australia.\\
 Current address: Centro de Radioastronom{\'i}a y Astrof{\'i}sica, UNAM,
 Campus Morelia, M{\'e}xico.
} 
\email{$^{\star}$p.bonfini@crya.unam.mx}

\author{Alister W.\ Graham}
\affil{Centre for Astrophysics and Supercomputing, Swinburne University of
  Technology, Victoria 3122, Australia.}

\begin{abstract}

 \noindent 
Partially-depleted cores are practically ubiquitous in luminous early-type
galaxies (M$_{B}\lesssim-$20.5 mag), and typically smaller than 1 kpc.  
In one popular scenario, supermassive black hole
binaries --- established during dry (i.e.\ gas-poor) galaxy mergers --- kick
out the stars from a galaxy's central region via three-body interactions.
Here, this ``binary black hole scouring scenario'' is probed at its extremes by
investigating the two galaxies reported to have the largest partially-depleted
cores found to date: 2MASX~J09194427+5622012 and 2MASX~J17222717+3207571 (the
brightest galaxy in Abell~2261).
We have fit these galaxy's two-dimensional light distribution using the
\corser{} model, and found that the former galaxy has a \corser{} break radius $R_{b,cS}=0.55$~kpc, three times smaller than the
published value.
We use this galaxy to caution that other reportedly large break radii
may too have
been over-estimated if they were derived using the ``sharp-transition'' (inner
core)-to-(outer S\'ersic) model.
In the case of 2MASX~J17222717+3207571, we obtain $R_{b,cS}=3.6$~kpc.
While we confirm that this is the biggest known partially-depleted
core of any galaxy, we stress that it is larger than expected from the evolution
of supermassive black hole binaries --- unless one invokes substantial
gravitational-wave-induced (black hole)-recoil events.
Given the presence of multiple nuclei located (in projection) within the core
radius of this galaxy, we explored and found support for the alternative
``stalled infalling perturber'' core-formation scenario, in which this galaxy's
core could have been excavated by the action of an infalling massive perturber.

\end{abstract} 




\section[Introduction]{Introduction}
\label{Introduction}

\noindent
The spheroidal component of 
many luminous early-type galaxies (ETGs; $M_{B} \lesssim -20.5\pm0.75$~mag)
are characterized by the presence of a depleted stellar core, which
manifests itself as a marked flattening of the inner light distribution
relative to the inward extrapolation of the spheroid's outer \Sersic{} profile
\citep[e.g.][]{graham:corser}.
Galaxies with shallow inner surface brightness profiles have of course been observed for
decades \citep[e.g.][]{king:1966,king:1972} but it has not always been reliably
established if this represents a deficit of stars relative to the outer profile,
nor have the sizes of these cores been robustly measured.
It is important to realize that a flat core does not necessarily represent a
signature of a depleted core, and flat cores can even arise from the presence of
additional nuclear components
\citep[see][their Appendix~A.2]{dullo:2012,dullo:2013}.
Furthermore, several brightest cluster galaxies \citep[BCGs; e.g.][]{donzelli},
and even some cD galaxies embedded in halos of intracluster light
\citep[NGC~4874 and UGC~9799;][]{seigar}, are characterized by low \Sersic{}
index spheroids whose flat inner light profiles do not deviate from their outer
\Sersic{} profile.

In \cite{Holm15A} we demonstrated that the galaxy Holm~15A, alleged to
have the largest depleted core on record \citep[4.6~kpc][]{lopez}, actually has
no central deficit relative to its outer \Sersic{} profile which also describes
the inner profile.
This is important because the original investigation led to \emph{(a)} the claim
of a massive depleted core and in turn \emph{(b)} the possible presence of a
truly massive $10^{11}~M_{\odot}$ black hole.
Remarkably, our results were later independently confirmed by the study of
\cite{madrid:2016}, who also described Holm~15A as ``core-less''.

The same situation may well have occurred with the analysis of a similar-looking
light profile, from the central galaxy in the MS0735.6+7421 cluster, for which
\cite{mcnamara} report a depleted core with a radius of 3.8 kpc --- which they
wrote supported the evidence for an ultramassive black hole.
Compounding matters, the allegedly largest black hole mass directly measured via
dynamical methods, specifically \MBH{} = $1.7\pm0.3\times10^{10}~M_{\odot}$
in NGC~1277 \citep[]{vandenbosch}, was later reduced to $5\times10^9~M_{\odot}$
\citep{emsellem,walsh} and is likely less than $1.2\times10^9~M_{\odot}$
\citep{graham:2016}.
Furthermore, it appears that the masses of most AGNs may have been over-estimated
by a factor of a few, due to an incorrectly calibrated virial factor
\citep{shankar}.
Collectively, this casts some doubt on the abundance of ultramassive black
holes, and, more specifically, reveals the need for an independent confirmation
of large black hole masses and large partially depleted cores.
Here we investigate the next two largest cores reported in the literature.

One of the favored scenarios for the formation of depleted stellar cores involves
binary supermassive black holes
\citep[SMBHs;][]{begelman,ebisuzaki,quinlan,yu,merritt:2005}.
It has been suggested that SMBH binaries, established during dry
(i.e.\ gas-poor) mergers which built the ETG, remove stars on radial orbits
crossing the galaxy nucleus via three-body interactions.
This activity is performed at the expense of the potential energy of the binary, 
which ultimately coalesces into a central SMBH.
Observational evidence supporting this theory is provided by the existing
scaling relations between the mass of the SMBH and the characteristic radius of
the depleted core, or the depleted stellar mass
\citep[e.g.][]{graham:2004,ferrarese:2006,lauer:2007}.
However, predictions from the SMBH binary scenario cannot readily account for the
extremely large cores, and hence depleted masses, reported for several ETGs
\citep[i.e.\ core radius $>1$ kpc; e.g.][]{laine,lauer:2007,postman,hyde}, unless
it is assumed that the host galaxy underwent extraordinary merging activity
\citep[see the discussion in][]{Holm15A}.
Reports of such large cores is therefore casting shadows on the SMBH scouring
scenario as the sole mechanism for core formation.

Several alternative models for the formation of cores can not yet boast the same
observational support which the SMBH scouring scenario received over the last
few decades, but they are gaining attention due to their ability to reproduce
larger cores (which may or may not be real).
Among these alternatives, the most noticeable is (arguably) the ``ejected SMBH''
scenario \citep[e.g.][]{redmount,merritt:2004,boylan,gualandris}.
According to this framework, the coalescence of the SMBH binary causes the newly
formed SMBH to recoil following the anisotropic emission of gravitational
waves in the other direction \citep{bekenstein,fitchett,gonzales}.
Potentially, the SMBH may be expelled from the galaxy core, or placed on a
radial orbit recurrently intersecting with the nucleus.
This phenomenon is able to significantly enlarge the core produced during
the SMBH binary phase
\citep[up to $\sim$5\% of the galaxy half-light radius;][]{gualandris}.
Arguments against this model have been raised in regard to the
``final parsec problem'', i.e.\ the possible stalling of the SMBH binary
separation (preventing the final coalescence) due to the depletion of core
stars capable of transferring momentum \citep{final_parsec}.
However, recent works based on realistic galaxy potentials considering triaxiality,
asymmetry, eccentricity of the SMBHs orbits, or rotation have shown that
the binary evolution is convergent
\citep[e.g.][]{khan,merritt:2015,holley,vasiliev}.
The SMBH recoil scenario has been recently advocated by \cite{markakis}
to explain the $\sim$0.2~kpc core they observed in the peculiar galaxy
NGC~3718. 

Another promising model for the explanation of large cores is the
``stalled perturber'' scenario.
In a seminal paper, \cite{chandrasekar} suggested that the dynamical friction exerted by
a homogeneous mass distribution can cause a captured object to spirally infall
due to the transfer of angular momentum from the infalling object to the background
particles (stars), which are moved to larger orbits.
More recently, \cite{read:2006a} reviewed the assumption of homogeneous mass
distribution, and demonstrated instead that a necessary condition for the spiral infall
is that the background particles cannot have a constant density distribution.
In particular, numerical simulations have shown that the infall of a clumpy baryonic
perturber makes a central ``cuspy'' dark matter distribution shallower, and can even
convert it to a constant density core
\citep[e.g.][]{elzant:2001,merritt:2002,elzant:2004,merritt:stalled_binary,tonini}.
\cite{goerdt} suggests that a baryonic/stellar core forms because the cusp is literally
shredded by the tidal interaction with the perturber \citep[see also][]{petts}. 

Apart from the ``ejected SMBH'' and the ``stalled infalling perturber'' models
considered above, other models allow for extremely large cores/mass deficits, such
as the ``multiple-SMBH scouring'' scenario of \cite{kulkarni}, or the combined
``sinking SMBH -- AGN feedback'' scenario of \cite{martizzi}.
However, the predictions from those models require much more tuning, and they
are somewhat difficult to explore observationally.

It is important to confirm claims of unusually large cores in order to: \emph{a})
check on the need for scenarios proposed to explain them, and \emph{b}) constrain
the existing scaling relations between the characteristics of the core and the mass
of the black hole \citep[\MBH{}; e.g.][]{rusli,dullo:2014}.
Massive ETGs are expected to host the most massive SMBHs and have the widest cores.
Although larger cores should be the easiest to measure, our recent study of 
Holm~15A revealed that misinterpretations are still possible.
These considerations motivated us to revisit the large cores reported in two galaxies:
2MASX~J17222717+3207571, the BCG of Abell~2261
\citep[hereafter A2261-BCG; $R_{core}$ = 3.2~kpc;][]{postman},
and 2MASX~J09194427+5622012
\citep[or SDSS-J091944.2+562201.1, hereafter SDSS-H5;\footnote{
 Our naming simply follows the indexing of the sample in \cite{hyde}.
} $R_{core}$ = 1.6~kpc;][]{hyde}.
Details on these galaxies are reported in Table~\ref{table:sample}.

\renewcommand{\tabcolsep}{1.0em}

\begin{deluxetable*}{lcccccccc}
 \tabletypesize{\small}
 \tablecaption{Sample\label{table:sample}}
 \tablehead{
  \colhead{Target}         &
  \colhead{$z$}            &
  \colhead{$D_{L}$}        &
  \colhead{$m-M$}          &
  \colhead{Scale}          &
  \colhead{Camera/Filter}  &
  \colhead{Exposure}       &
  \colhead{Pixel Scale} 
  \\
  \colhead{}                  &
  \colhead{}                  &
  \colhead{[Mpc]}             &
  \colhead{[mag]}             &
  \colhead{[kpc/$\arcsec$]}   &
  \colhead{}                  &
  \colhead{[sec]}             &
  \colhead{[$\arcsec$/pixel]} 
  \\
  \colhead{{\tiny (1)}} &
  \colhead{{\tiny (2)}} &     
  \colhead{{\tiny (3)}} &
  \colhead{{\tiny (4)}} &
  \colhead{{\tiny (5)}} &
  \colhead{{\tiny (6)}} &     
  \colhead{{\tiny (7)}} &
  \colhead{{\tiny (8)}}
 }
 \startdata
2MASX~J17222717+3207571 (A2261-BCG) & 0.225 & 1119 & 40.24 & 3.61 & ACS/F814W  & 4099 & 0.050 \\
\addlinespace 
2MASX~J09194427+5622012 (SDSS-H5)   & 0.278 & 1423 & 40.77 & 4.22 & HRC/F775W & 1200 & 0.025 \\
 \enddata
 \tablecomments{
  Basic information for the sample galaxies, and for the HST images used in the current work.
  \\
  $^{(1)}$ Target name.
  $^{(2)}$ Redshift measurement from SDSS-DR1 \citep{SDSS:DR1} for SDSS-H5,
           and from NED (Virgo + GA + Shapley) for A2261-BCG.
  $^{(3)}$ Luminosity distance, assuming a cosmology
           with $H_{0}$ = 70~km s$^{-1}$ Mpc$^{-1}$, $\Omega_{\Lambda}$ = 0.7, and $
           \Omega_{m}$ = 0.3.
  $^{(4)}$ Distance modulus.
  $^{(5)}$ Scale at the luminosity distance of column 3.
  $^{(6)}$ HST camera and filter.
  $^{(7)}$ Total exposure time.
  $^{(8)}$ Instrument pixel scale.
 }
\end{deluxetable*}

As warned in \cite{graham:corser}, the Nuker model \citep{lauer:nuker} can
incorrectly imply the presence of a partially depleted core when there is an
un-disturbed (non-depleted) \Sersic{} profile with a low \Sersic{} index.
This occurred in --- for example --- NGC 4473 (\citealt{pinkney};
see \citealt{dullo:2014}) and Holm~15A (\citealt{lopez}; see \citealt{Holm15A}).
As was also explained in \cite{graham:corser}, the Nuker model ``break radius''
can significantly over-estimate the sizes of cores (see \citealt{trujillo:corser}
and \citealt{dullo:2012,dullo:2014}) which led to the alternative use of the radius where the
negative logarithmic slope of the radial intensity profile equals 0.5.
However every light profile has such a radius, irrespective of whether or not
it actually contains a partially depleted core.
It is therefore necessary to test if there is an inner deficit of light relative
to the outer light profile.
In this work we use the \corser{} model to do this.
While A2261-BCG has so far only been fit with a Nuker model, SDSS-H5 has already
been fit with a \corser{} model by \cite{hyde}.
However, we identified some concerns with their approach (see \S\ref{Discussion})
and we therefore perform an independent fit analysis, finding a core which is
three times smaller for the reasons explained within.

This paper is structured as follows.
In \S\ref{Data}, we present the data.
In \S\ref{Modelling} we outline our strategy for the 2D fit of the surface
brightness distribution, while in \S\ref{Results} we present the results of this
analysis.
In \S\ref{Discussion} we discuss our results in the context of different formation
scenarios for (massive) depleted cores.
Finally, we summarize our conclusions in \S\ref{Conclusions}.
Throughout the paper, we assume a cosmology with $H_{0}$ = 70~km s$^{-1}$ Mpc$^{-1}$,
$\Omega_{\Lambda}$ = 0.7, and $\Omega_{m}$ = 0.3.

\section[Data]{Data}
\label{Data}

\noindent
Our analysis is performed on archival $HST$ images obtained in the $I$-band.
This provided the best compromise between high spatial resolution and minimal dust
contamination\footnote{
 We performed a visual inspection of the F225W (WFC3/UVIS), F475W (ACS), and
 F606W (ACS) images for A2261-BCG, and of the F475W (ACS) image for SDSS-H5, and
 found no obvious trace of dust contamination, although it is not excluded that
 --- given the large distance of the galaxies --- unresolved dust lanes might
 still be present.
}, both desirable in the study of galaxy cores.
In particular, we retrieved deep ACS/F814W (Johnson-Cousins $I$) and
ACS-HRC/F775W (SDSS $i$) images for A2261-BCG and SDSS-H5, respectively, from the
\textsc{STScI MAST} Archive (see Table~\ref{table:sample} for image specifics).
The same camera/filter sets were also used by \cite{postman} and \cite{hyde},
therefore allowing for a direct comparison.
For each galaxy, we combined the different exposures using the
\textsc{AstroDrizzle} tool \citep[v1.1.16;][]{AstroDrizzle} through the
PyRAF (v2.1.6) suite\footnote{
 PyRAF is a product of the Space Telescope Science Institute, which is operated
 by AURA for NASA.
}, sampling the images at the native pixel scale of each camera (see Table
\ref{table:sample}), which was more than sufficient to study the cores of our
sample galaxies.

\subsection[Masking]{Masking}
\label{Masking}

\begin{figure*}

 \definecolor{light-gray}{gray}{0.45}
 \linethickness{0.5pt}

 \makebox[\linewidth]{
  \begin{overpic}[width=0.48\textwidth]
   {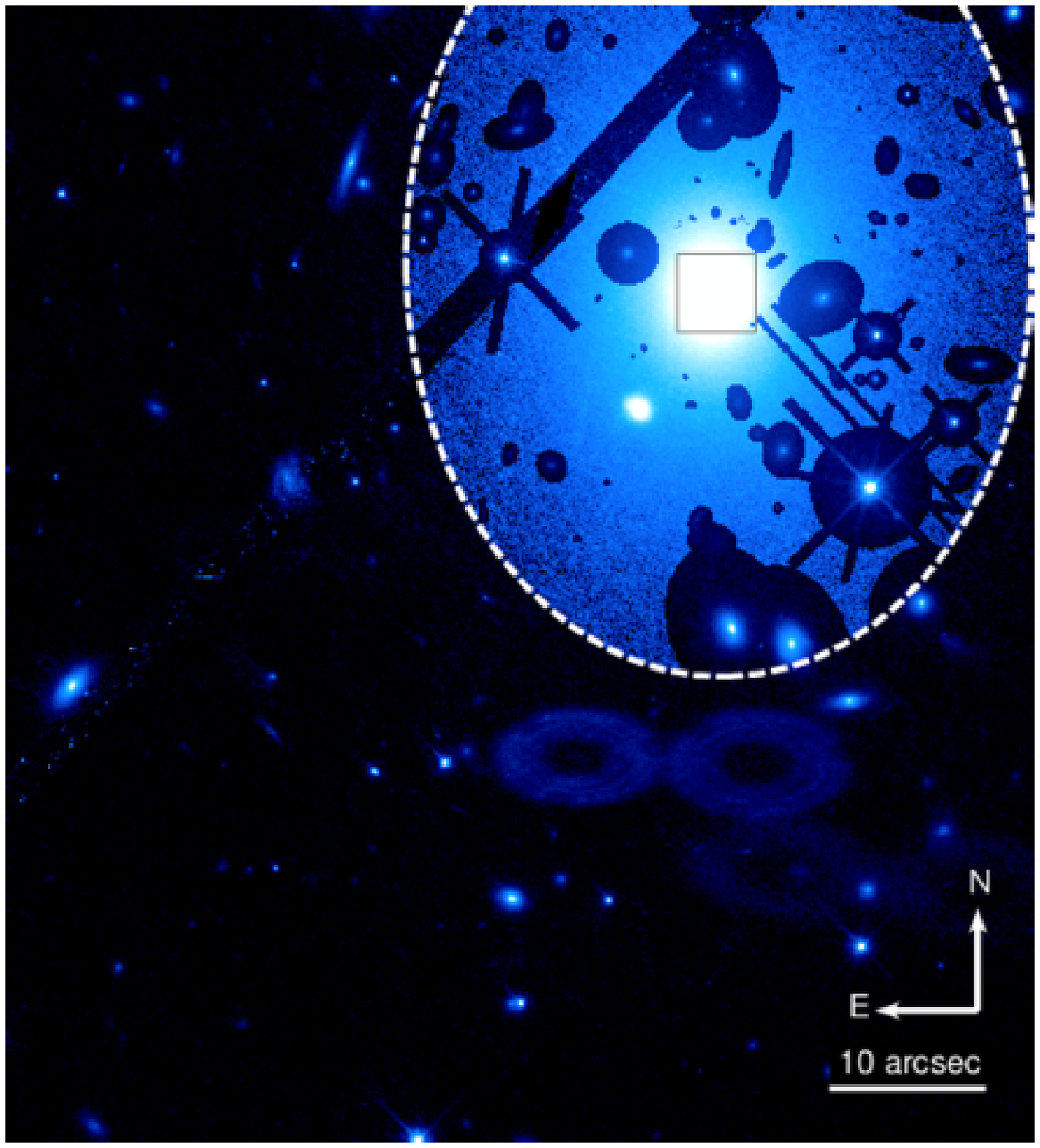}
   \put(5,5){\includegraphics[scale=0.213]{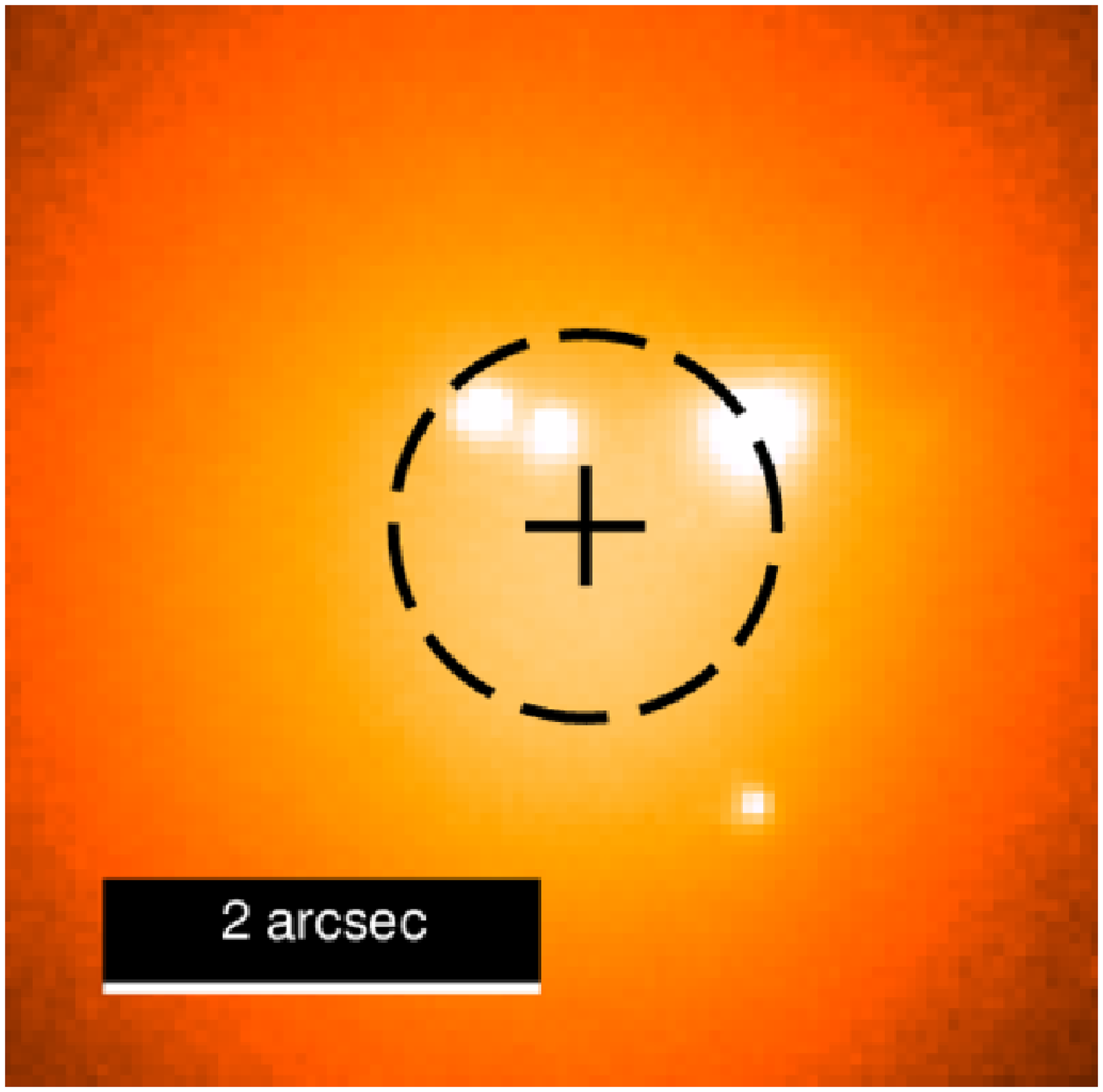}}
   \put(5,90){\textcolor{white}{A2261-BCG}}
   \put(12,40){\textcolor{black}{A2261-BCG core}}
   \put(20,33){\textcolor{black}{1}}
   \put(26,28){\textcolor{black}{2}}
   \put(31,31){\textcolor{black}{3}}
   \put(32,17){\textcolor{black}{4}}
   \put(35,29){\textcolor{black}{5}}
   \put(06,43.5){\color{light-gray}\line(5,3.2){52}}
   \put(44,06)  {\color{light-gray}\line(1.58,4.8){21.1}}
  \end{overpic}

  \begin{overpic}[width=0.48\textwidth]
   {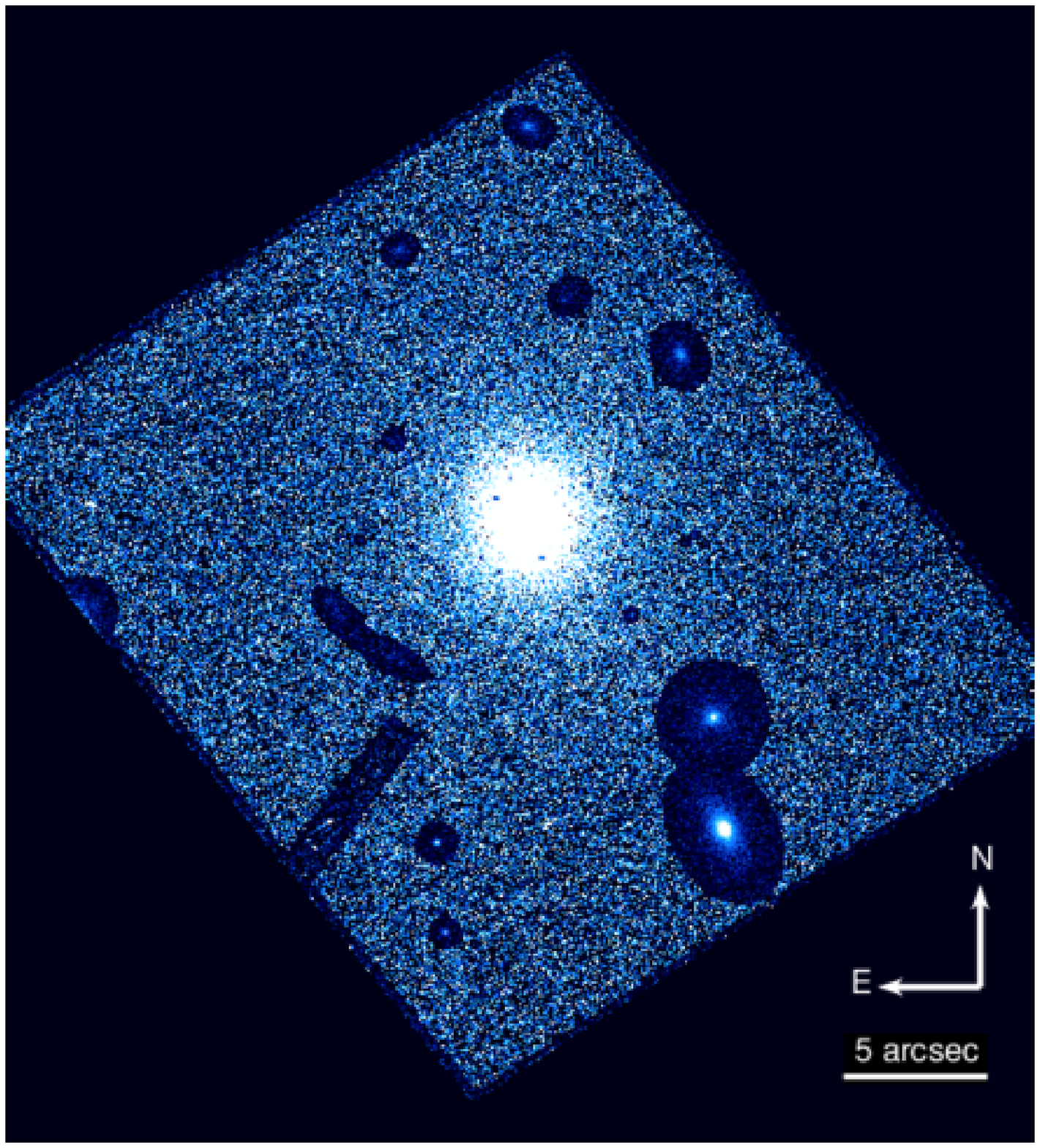}
   \put(5,90){\textcolor{white}{SDSS-H5}}
  \end{overpic}
 }
 \makebox[\linewidth]{
  \begin{overpic}[width=0.48\textwidth]
   {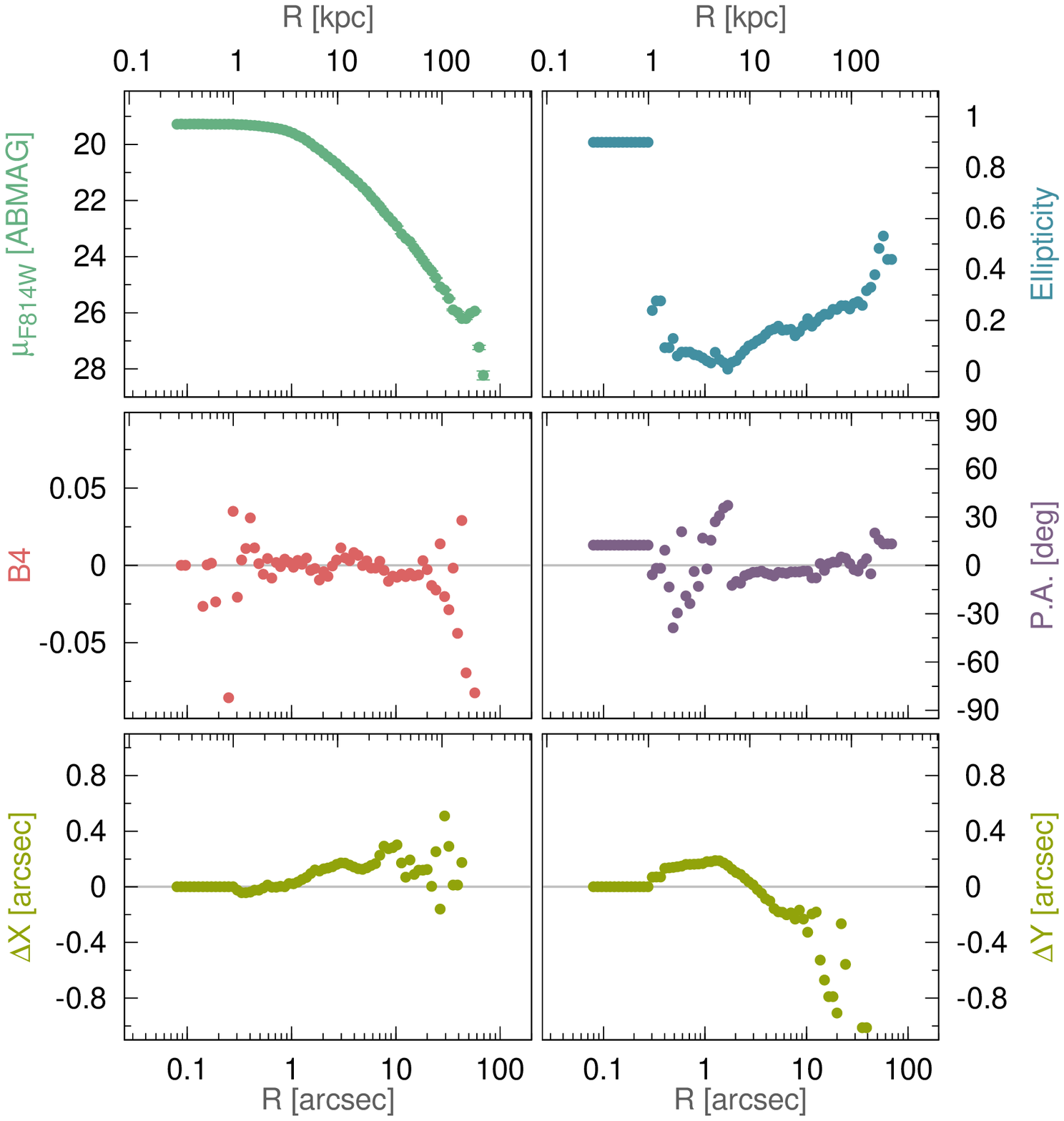}
  \end{overpic}

  \begin{overpic}[width=0.48\textwidth]
   {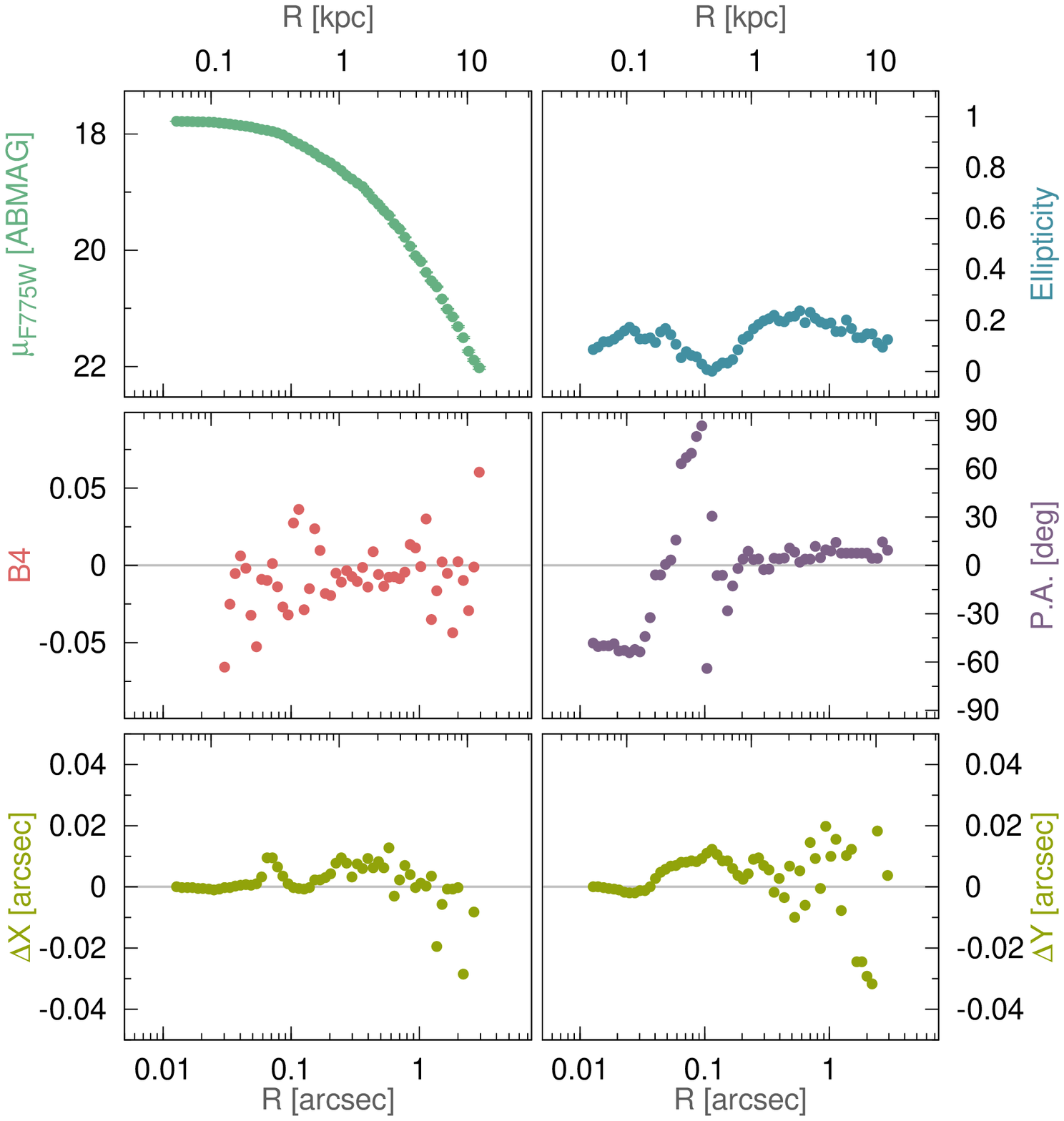}
  \end{overpic}
 }
 \caption{
  \emph{HST} images of the sample galaxies, and preliminary 1D analysis for
  A2261-BCG (\emph{left}) and SDSS-H5 (\emph{right}).
  \newline
  \emph{Top.-}
  \emph{HST} mosaics produced with the procedure described in
  \S\ref{Data}.
  The darker regions correspond the areas excluded (masked) from our analysis
  (see \S\ref{Masking}).
  The dashed white ellipse in the top-left panel corresponds to the physical extent
  of our 2D fit (see \S\ref{Modelling}).
  The insert in the top-left panel is a zoom into the central regions of A2261-BCG.
  The dashed black circle represents the size of the core as measured
  by \cite{postman}, and its center (black cross) corresponds to the centroid of
  the innermost elliptical isophote found with our IRAF.\emph{ellipse} analysis
  (see \S\ref{Preliminary photometry}).
  The objects (1 -- 4) around the core have been labelled following \cite{postman}.
  After subtracting the modelled galaxy light, we revealed an additional object (5),
  not visible in this representation.
  \emph{Bottom.-}
  IRAF.\emph{ellipse} major-axis radial profiles for: surface brightness
  (\emph{top-left}), ellipticity (\emph{top-right}), 4\emph{th} harmonic deviation
  from perfect ellipticity ($B4$; \emph{middle-left}), position angle
  (\emph{middle-right}), and isophote centroid shift along the $x$-axis
  (\emph{bottom-left}) and $y$-axis (\emph{bottom-right}) with respect to the
  innermost isophote.
  The photometric errors of the surface bright ness data points are typically within
  the size of the data symbols.
  \label{figure:mosaic_and_ellipse}
 }
\end{figure*}

\noindent
We masked the contaminating objects in the field using the detections obtained from
a double run of \textsc{SExtractor} \citep{SExtractor}, tuned first to identify
point-like sources and then extended objects.
We then additionally masked cosmic rays, chip imperfections, and internal
reflections after visual inspection.
The masked areas are shown in the images of Figure~\ref{figure:mosaic_and_ellipse}
(top-panels) as darkened regions.

Notice that we did not mask a bright object S--E of the center of A2261-BCG;
given the significant overlap/contamination, we preferred instead to
model it concurrently with A2261-BCG.
Similarly, we did not mask the bright ``knots'' around the core of this galaxy
(Figure~\ref{figure:mosaic_and_ellipse}, inset of the top-left panel)
but instead modelled them.
In their analysis, \cite{postman} labelled these knots with progressive numbers
from 1 to 4 (in clockwise order in the image), a nomenclature which we retain here
for the sake of comparison.
However, from our initial 2D fit residuals, we identified a fifth knot (``knot 5''),
not directly visible in this image.
Knots 1---5 have been included as additional components in our fit, similarly to
the bright object S--E of the center.
The nature of these knots will be discussed in \S\ref{Discussion}.

\subsection[PSF and sigma image]{PSF and sigma image}
\label{PSF and sigma image}

\noindent
Our 2D fitting algorithm \CORSAIR{} (\citealt{corsair}; see also \S\ref{Modelling})
convolves, at each iteration, the model with the point spread function (PSF), and
then compares the result against the data.
To create realistic PSFs for our images, we first used the TinyTim tool
\citep{TinyTim} to produce ``distorted'' $HST$ PSFs, i.e.\
as they would appear in the native $HST$ fields (without applying the HST
distortion geometry corrections).
We then run these PSF images through \textsc{AstroDrizzle} using the exact same
setup used to create the mosaics.
In this way we obtained artificial PSFs whose features (e.g.\ diffraction spikes)
are oriented exactly as those of the point-like sources in the mosaics.
Moreover, this process allowed us to to mimic the ``pixellation'' imprint which \textsc{AstroDrizzle} produced on the mosaicked images.

\textsc{AstroDrizzle} can provide variance maps as an additional output of the
mosaicing process. 
Apart from the Poissonian errors on the source and background fluxes, these
variance images include the flat-fielding uncertainties and the correlated pixel
noise.
However, since in the mosaics of both galaxies the galaxy light contaminates the
sky background (see \S\ref{Preliminary photometry}), which is by far the main
responsible for the "error budget", we preferred not trust the internal
\textsc{AstroDrizzle} algorithm for the sky estimation, and decided to generate
our own weight images.
The weight (or ``sigma'') image for SDSS-H5 was created using the \CORSAIR{}
algorithm (inherited from \GALFIT{}), which uses a sigma-clipping
technique to derive the image noise characteristics.
Instead, for the sigma image of A2261-BCG --- whose field is significantly crowded ---
we preferred to proceed following the procedure described in
\citet[their Section 3.2.1]{corsair}.
In brief, we performed the sum in quadrature of the Poissonian noise of the source
counts plus the background noise.
The background noise is in turn the sum in quadrature of the noise due to the
readout, the detector dark current, and the sky.
While the first two are known (since they are characteristics of the camera), the
latter had to be estimated.
The WFC3 Instrument Handbook \citep[][their Table 9.7.1]{WFC3} provides rough
estimates for the surface brightness of the zodiacal light (the dominant component
of sky background in typical observing conditions) in the $V$-band, as a function
of elliptical coordinates.
We used the astrometric information in our images to pick the relevant $V$-band sky
magnitude from these values, and converted them to $I$-band
(as we were using the F814W and F775W filters) using an average sky color
\mbox{$V$ -- $I$ $\sim$ 0.6~mag} \citep[e.g.][]{vaccari}.
This provided us with an independent estimate for the sky magnitudes, which were
converted into counts using the exposure information.
Finally, the sky noise was calculated applying Poissonian statistics over these
expected sky counts.

\section[Modelling]{Modelling}
\label{Modelling}

\noindent
We performed two-dimensional fits to the surface brightness distribution of the
galaxies using \CORSAIR{}\footnote{
 \url{www.astronomy.swin.edu.au/~pbonfini/galfit-corsair/}
} \citep{corsair}, an advancement of the \textsc{Galfit} software \citep{GALFIT}
which we developed to include the \corser{} model \citep{graham:corser}.
The 2D approach has the advantage of being able to simultaneously fit (rather
than mask) contaminant objects.
This ability is particularly handy for the case of A2261-BCG, since several
objects lie (along the line of sight) within its central regions
(see Figure~\ref{figure:mosaic_and_ellipse}, top-left panel), and they could
affect the estimate of, and be related to, the core size.
In our recent work \citep[][]{corsair,Holm15A}, we showed that 2D
fitting proved to agree with 1D analysis modulo strong radial ellipticity
gradients (which are of minor concern in the present case; see the ellipticity
profiles in Figure~\ref{figure:mosaic_and_ellipse}, and
\S\ref{Preliminary photometry} for how they were measured).

\subsection[Preliminary photometry]{Preliminary photometry}
\label{Preliminary photometry}

\noindent
When creating the mosaics, we used the default \textsc{AstroDrizzle} setup except
for the automatic sigma-clipping sky subtraction, due to the relatively large
extent on the chip of the galaxy (in the case of SDSS-H5) and for the high source
density around A2261-BCG.
Instead, we evaluated the sky-background by hand, measuring it at the edges of
the images over several ``boxes'' located at different azimuthal angles around each galaxy, and then adopting the median value, which was then subtracted from the
images.
The boxes were located at galactocentric distances of $\sim$10 $R_e$ and
$\sim$4 $R_e$, for A2261-BCG and SDSS-H5 respectively.
Therefore, especially for the case of SDSS-H5, we could not exclude that our
assumed background was contaminated by the galaxian light.
The influence of this uncertainty on our best-fit parameters has been quantified
via the Monte Carlo simulation presented in the Appendix. 

The ``first-guess'' parameters for the models have been chosen after inspecting
the various 1D profiles, measured along elliptical isophotes
using the IRAF.\emph{ellipse} task \citep{ellipse}.
While the IRAF.\emph{isofit} task \citep{ciambur} is superior, it produces similar
results when dealing with rather round ETGs that do not contain edge-on disks.
The bottom panels of Figure~\ref{figure:mosaic_and_ellipse} show the surface
brightness ($\mu$) of the galaxies, extracted along the semi-major axis, along with
the radial profiles of: ellipticity ($e$); 4\emph{th} harmonic deviation
from perfect ellipticity ($B4$ "boxiness/diskiness" parameter); position angle (P.A.);
and isophote centroid shift.

In running IRAF.\emph{ellipse} on the image of A2261-BCG, we excluded the knots
1--5 and the overlapping companion galaxy, which were included in our 2D fit
(see \S\ref{Masking} and Figure~\ref{figure:projection}).
Due to the large fraction of masked pixels and the
relatively mild slope of the
brightness profile in the core region, the centering algorithm of
IRAF.\emph{ellipse} failed within the innermost $\sim$3\arcsec of A2261-BCG.
The data points within this limit are therefore measured along concentric ellipses
of identical ellipticity, P.A., and center.
However, as already noticed in \citet[their Figure 4]{postman}, a contour plot
of the galaxy reveals that, within the core, the center of the isophotes are
slightly shifted N--W by $\sim$1$\arcsec$ with respect to the outer isophotes
(see \S\ref{Fitting}).
This feature will be addressed further in \S\ref{Discussion}.

\subsection[Fitting]{Fitting}
\label{Fitting}

We fit the surface brightness distribution of the galaxies using a 2D
\corser{} model \citep{graham:corser}, which provides a smooth connection
between an outer \Sersic{} component and an inner power-law component.
The \corser{} profile can be expressed as:

\begin{small}
\begin{equation}
 I(R) = I'\left[1+\left({R_{b,cS} \over R}\right)^{\alpha}\right]^{\gamma / \alpha}
        \exp \left[ -b_{n}\left({ {R^{\alpha} + R_{b}^{\alpha}} \over R_{e}^{\alpha} }\right)^{{1 / (n\alpha) }}\right]
 \label{equation:corser}
\end{equation}
\end{small}  

\noindent
with

\begin{small}
\begin{equation}
 I' = I_{b} 2^{-\gamma/\alpha} \exp \left[b_{n} 2^{1 / n\alpha} (R_{b,cS}/R_{e})^{1/n}\right]
 \label{equation:IP}
\end{equation}
\end{small}  

\noindent
where $R_{b,cS}$ (\corser{} ``break radius'') is the radius corresponding to the
mid-point between the outer \Sersic{} and the inner power-law portion of the
profile. The parameter $\alpha$ modulates the sharpness of the changeover, and
$I_{b}$ represents the intensity at $R_{b,cS}$.
The asymptotic inner power law slope is given by $\gamma$, while the outer \Sersic{}
index is identified by $n$.
Finally, $b_{n}$ is a normalization factor which can be defined to make $R_{e}$ the
effective half-light radius of the non-depleted \Sersic{} profile
\citep[see][]{graham:corser}.
We remark that use of $\alpha$, when the transition between the core and outer region
is broad, is important for not over-estimating the size of the break radius. 

In this context, the break radius $R_{b,cS}$ is adopted as a parametric
measurement of the core size.
This is a formally defined point, and it should borne in mind that a SMBH binary
which excavates a core can affect a galaxy light profile up to radii well beyond
the break radius.
In fact, stars on extremely elongated (almost radial) orbits can intersect
with the sphere of influence of the SMBH binary, and hence be ejected.
It is of course possible to designate alternative ``core sizes'' defined
relatively to the deviation from the outer \Sersic{} fit.
However, any radius corresponding to a the deviation from the outer \Sersic{}
fit of, say, 0.1\%, or 0.5\%, or x\%, will occur far out in the profile for
finite values of alpha (since the \corser{} and its \Sersic{} part only converge for $R\rightarrow\infty$).
Moreover, any such radius is arbitrary and therefore not a radius that is important.
Notice that the physical quantity of actual interest in the study of depleted
cores --- the "luminosity (or mass) deficit" --- is independent, in our analysis,
from the definition of $R_{b,cS}$, being the difference between the integrated
luminosity (or mass) of the extrapolated \Sersic{} profile and that of the actual
\corser{} profile.

The fit area was extended to the whole image in the case of SDSS-H5, while it
was arbitrarily limited within the isophote with a semi-major axis of 27$\arcsec$.5
(corresponding to a surface brightness of 25.5~\magsb{} in the F814W filter)
for A2261-BCG.
The fit areas appear as brighter regions in the top panels of Figure
\ref{figure:mosaic_and_ellipse}.

For the knots within the core region of A2261-BCG, we initially assumed point-like
models (i.e.\ PSFs), and then iteratively increased the complexity of the
components after inspection of the residuals (first using Gaussian and then
\Sersic{} models).
In our final fit, knots 1--3 are described by \Sersic{} components, while knots
4 and 5 by point-like sources.
The companion galaxy S--E of A2261-BCG was fit with a Sersic{} model. 

Given the isophote centroid shift within the core region of A2261-BCG
(see Figure~\ref{figure:mosaic_and_ellipse}), we decided to perform an additional
fit limited to 5$\arcsec$ in order to allow a local re-centering of the underlying
\corser{} component.
This analysis obtained a better fit of the knots, rather than a refined estimate
of the core size (for which the measurement of the outer \Sersic{}
profile is fundamental; see \S\ref{Comparing our fit results with the literature}).

\renewcommand{\tabcolsep}{0.5em}
\def\arraystretch{1.5} 

\begin{turnpage}

\begin{deluxetable*}{lccccccccccccccccc}
 \tabletypesize{\small}
 \tablecaption{Best-fit \corser{} Parameters \label{table:corser}}
 \tablehead{
  \colhead{Target}                         &
  \colhead{Filter}                         &
  \colhead{$\mu_{b,filter}^{\dagger}$} &
  \colhead{$m_{filter}^{\dagger}$}      &
  \colhead{$R_{b,cS}$}                   &
  \colhead{$\alpha$}                    &
  \colhead{$\gamma$}                    &
  \colhead{$R_{e}$}                      &
  \colhead{$n$}                          &
  \colhead{$e$}                          &
  \colhead{$P.A.$}
  \\
  \colhead{}                            &
  \colhead{}                            &
  \colhead{[mag/$\arcsec^{2}$]}      &
  \colhead{[mag]}                       &
  \colhead{[$\arcsec$] / [kpc]}      &
  \colhead{}                            &
  \colhead{}                            &
  \colhead{[$\arcsec$] / [kpc]}      &
  \colhead{}                            &
  \colhead{}                            &
  \colhead{[deg]}
  \\
  \colhead{{\tiny (1)}}  &
  \colhead{{\tiny (2)}}  &     
  \colhead{{\tiny (3)}}  &
  \colhead{{\tiny (4)}}  &
  \colhead{{\tiny (5)}}  &
  \colhead{{\tiny (6)}}  &     
  \colhead{{\tiny (7)}}  &
  \colhead{{\tiny (8)}}  &
  \colhead{{\tiny (9)}}  &
  \colhead{{\tiny (10)}} &
  \colhead{{\tiny (11)}}
 }
 \startdata
A2261-BCG & F814W &
	                  19.59$^{+0.04}_{-0.00}$ &
			  14.77$^{+0.00}_{-0.21}$ &
			  1.01$^{+0.96}_{-0.01}$ /
			  3.63$^{+0.96}_{-0.01}$ &
			  3.6$^{{\tiny \ ^{\vee}}}_{-0.8}$ &
			  0.02$^{{\tiny \ ^{\vee}}}_{-0.04}$ &
			  10.77$^{+70.42}_{{\tiny \ _{\wedge}}}$ /
			  38.86$^{+70.42}_{{\tiny \ _{\wedge}}}$ &
			  3.9$^{+1.3}_{-0.0}$ &
			  0.17$^{+0.00}_{{\tiny \ _{\wedge}}}$ &
			  0.3$^{+0.1}_{-0.0}$ & 
			  \\ 
\addlinespace 
\addlinespace 
SDSS-H5 & F775W &
	                  18.04$^{+0.06}_{-0.01}$ &
			  16.32$^{+0.05}_{-0.22}$ &
			  0.13$^{+0.60}_{-0.08}$ /
			  0.55$^{+0.60}_{-0.08}$ &
			  1.2$^{+0.7}_{-0.2}$ &
			  0.07$^{+0.32}_{-0.11}$ &
			  4.65$^{+57.66}_{-5.88}$ /
			  19.63$^{+57.66}_{-5.88}$ &
			  5.2$^{+1.8}_{-0.5}$ &
			  0.18$^{+0.00}_{{\tiny \ _{\wedge}}}$ &
			  7.2$^{+0.0}_{-0.0}$ & 
			  \\ 
\addlinespace 
\addlinespace 
  \addlinespace 

  \hline

  \addlinespace 
  \addlinespace 
  \addlinespace 
  \addlinespace 
  
  \hline
  \hline

  \addlinespace 
  
  {$m_{filter,corr}^{\dagger}$}      &
  {M$_{filter,0}^{\dagger}$}         &
  {$M/L$}                             &
  {\Msph{}}                            &
  {$M_{def}$}                         &
  {$\frac{M_{def}}{M_{\rm sph,*}}$} &
  {$\frac{M_{def}}{M_{\bullet}}$}
  \\
  {[mag]}                       &
  {[mag]}                       &
  {[$M_{\odot}/L_{\odot}$]} &
  {[$M_{\odot}$]}            &
  {[$M_{\odot}$]}            &
  {[\%]}                       &
  {}
  \\
  {{\tiny (12)}} &
  {{\tiny (13)}} &
  {{\tiny (14)}} &
  {{\tiny (15)}} &
  {{\tiny (16)}} &
  {{\tiny (17)}} &
  {{\tiny (18)}}
  \\

  \addlinespace 

  \hline

  \addlinespace 

14.55$^{+0.00}_{-0.21}$ &
	                  -25.69$^{+0.00}_{-0.21}$ &
			  3.5 &
			  4.44$^{+0.96}_{{\tiny \ _{\wedge}}} \times 10^{12}$&
			  1.75$^{+1.31}_{-0.91} \times 10^{11}$ &
			  3.9$^{+1.80}_{-2.00}$ &
			  6.9$^{+30.3}_{+1.3}$ \\ 
\addlinespace 
\addlinespace 
16.03$^{+0.05}_{-0.22}$ &
	                  -24.75$^{+0.05}_{-0.22}$ &
			  3.5 &
			  1.80$^{+0.41}_{-0.09} \times 10^{12}$&
			  0.81$^{+0.82}_{-0.40} \times 10^{11}$ &
			  4.5$^{+3.00}_{-2.00}$ &
			  7.7$^{+37.9}_{+1.6}$ \\ 
\addlinespace 
\addlinespace 
 \enddata
  
 \tablecomments{
  Results of the \corser{} fit to the sample galaxies.
  \\
  $^{(1)}$  Target name.
  $^{(2)}$  Reference filter for the magnitudes and surface brightnesses listed here.
  $^{(3)}$  Surface brightness at the \corser{} break radius.
  $^{(4)}$  Integrated apparent magnitude of the \corser{} model.
  $^{(5)}$  Break radius in units of arcseconds and kiloparsecs (the physical scale
              is provided in Table \ref{table:sample}).
  $^{(6)}$  Alpha parameter for the \corser{} model.
  $^{(7)}$  Inner power-law index for the \corser{} model.
  $^{(8)}$  Effective radius ($R_{e}$) of the \Sersic{} portion of the \corser{}
              profile, in units of arcseconds and kiloparsecs.
  $^{(9)}$ \Sersic{} index.
  $^{(10)}$ Model ellipticity.
  $^{(11)}$ Model position angle (\mbox{North = 0$^{\circ}$}).
  $^{(12)}$ Extinction and K-dimming corrected apparent magnitude.
              A galactic extinction of $\sim$0.07~mag (A2261-BCG) and
	      0.06~mag (SDSS-H5) was obtained from NED.
              We applied a $K$-correction of 0.15~mag (A2261-BCG) 
	      and 0.23~mag (SDSS-H5) following the prescriptions of
	      \cite{K-corr} for early-type galaxies.
	      In \cite{K-corr} and NED, we adopted the values for the Johnson-Cousins
	      $I$ and SDSS $i$ bands as proxies for those of the $HST$
	      $F814W$ and $F775W$ filters, respectively.
  $^{(13)}$ Absolute rest-frame magnitude of the object, adjusted for extinction
              and K-correction.
	      Redshift dimming has been accounted for in the luminosity
              distances which we used to calculate the distance modulus
              (provided in Table \ref{table:sample}).
  $^{(14)}$ Mass-to-light ratio ($M/L$; see \S\ref{Results} for details).
  $^{(15)}$ Model mass calculated assuming the $M/L$ of column 14. 
  $^{(16)}$ Mass deficit calculated from the luminosity deficit ($L_{def}$)
              assuming the $M/L$ reported in column 14;
              $L_{def}$ is in turn defined as the difference between the
	      integrated luminosity of the extrapolation of the \Sersic{} part
	      of the \corser{} model, and the luminosity of the \corser{} model
	      itself.
  $^{(17)}$ Relative mass deficit.
  $^{(18)}$ Ratio of mass deficit to SMBH mass (from the \MBH{} --$M_{\rm sph,*}$ relation,
              see Table~\ref{table:BH}).
  The errors reported in this table, which represent 50\% confidence levels,
  reflect the uncertainty on the sky background (which dominates the error budget),
  and have been estimated using the Monte Carlo simulation presented in the Appendix.
  The $\vee$ and $\wedge$ symbols represent upper and lower limits, respectively.
 }
 \tablenotetext{$\dagger$}{
  Values refer to the AB mag system.
  The zero-point for the calibration is provided in the $HST$ image header.
 }
\end{deluxetable*}

\end{turnpage}


\renewcommand{\tabcolsep}{0.5em}

\begin{deluxetable}{lccccc}
 \tabletypesize{\small}
 \tablecaption{Parameters for Knots 1--5 (A2261-BCG)\label{table:knots}}
 \tablehead{
  \colhead{Object}                        &
  \colhead{Model}                         &
  \colhead{$m_{F814W}^{\dagger}$}      &
  \colhead{$m_{F814W,corr}^{\dagger}$} &
  \colhead{M$_{F814W,0}^{\dagger}$}    &
  \colhead{$M_{*}$}
  \\
  \colhead{}                       &
  \colhead{}                       &
  \colhead{[mag]}                  &
  \colhead{[mag]}                  &
  \colhead{[mag]}                  &
  \colhead{[$M_{\odot}$]}
  \\
  \colhead{{\tiny (1)}}  &
  \colhead{{\tiny (2)}}  &     
  \colhead{{\tiny (3)}}  &
  \colhead{{\tiny (4)}}  &
  \colhead{{\tiny (5)}}  &
  \colhead{{\tiny (6)}}
 }
 \startdata
Knot 1 & \Sersic{} & 21.76 & 21.54 & -18.70 & 7.09e+09 \\ 
\addlinespace 
Knot 2 & \Sersic{} & 22.14 & 21.92 & -18.32 & 5.00e+09 \\ 
\addlinespace 
Knot 3 & \Sersic{} & 19.75 & 19.53 & -20.71 & 4.52e+10 \\ 
\addlinespace 
Knot 4 & PSF    & 23.22 & 23.00 & -17.24 & 1.86e+09 \\ 
\addlinespace 
Knot 5 & PSF    & 25.20 & 24.98 & -15.26 & 2.98e+08 \\ 
\addlinespace 
 \enddata
 \tablecomments{
  Results from simultaneously modelling the knots 1--5 and the inner 5$\arcsec$ (square).
  \\
  $^{(1)}$ Target name.
  $^{(2)}$ Model.
  $^{(3)}$ Integrated apparent magnitude.
  $^{(4)}$ Extinction and K-corrected apparent magnitude.
             (see description in \S\ref{Fitting}).
  $^{(5)}$ Absolute rest-frame magnitude of the object, adjusted for extinction
             and K-correction
	     (distance modulus is provided in Table \ref{table:sample}).
  $^{(6)}$ Stellar mass calculated assuming the same mass-to-light ratio
             used for A2261-BCG
             (3.5~$M_{\odot}/L_{\odot}$; see \S\ref{Results}).
 }
 \tablenotetext{$\dagger$}{
  Values refer to the AB mag system.
  The zero-point for the calibration is provided in the $HST$ image header.
 }
\end{deluxetable}



\section[Results]{Results}
\label{Results}

\noindent
Our measured \corser{} parameters for the galaxies are reported in 
Table~\ref{table:corser}, while the results from the modelling of the knots of A2261-BCG
are presented in Table~\ref{table:knots}.
In the same tables, we also report the stellar masses of each galaxy's spheroid
($M_{\rm sph,*}$) derived from the integrated stellar
luminosities assuming the stellar mass-to-light ($M/L$) ratio given in Table~\ref{table:corser}.
This $M/L$ was estimated using the ``Worthey model interpolation engine'' applet\footnote{
  \url{http://astro.wsu.edu/worthey/dial/dial_a_model.html}
}
based on the evolutionary models by \cite{M_L}, where we adopted the default
(Salpeter) initial mass function prescriptions, assuming a uniformly old
(12~Gyrs), solar metallicity stellar population (as typical for massive
early-type galaxies, e.g. \citealt{ETG_age_metallicity}).
The errors on the best-fit parameters reported in Table \ref{table:corser},
which represent 50\% confidence levels, are mostly due to the uncertainty
on the background level (which largely dominates over the Poissonian uncertainty
obtained by reversing the best-fit covariance matrix).
These errors were estimated by simulating different sky levels for the images,
and then repeating the same fits in order to evaluate the possible range
for each parameter.
The details of the estimation of the parameter errors are given in the Appendix.

In Figure~\ref{figure:residuals} we present the fit residuals
(image $-$ model) obtained with \CORSAIR{} for A2261-BCG (left) and
SDSS-H5 (right).
The insert in the left panel shows the residuals for the fit of the 5$\arcsec$
core region of A2261-BCG.
\newline
The top panels of Figure~\ref{figure:projection} present the 1D projections of the
surface brightness of the models (blue solid curve) measured along the same
galaxy isophotes identified with our IRAF.\emph{ellipse} analysis of each image
(see \S\ref{Preliminary photometry}).
It should be noted that these 1D projections are not the fit, which was performed
in 2D, but simply show the match along this single profile.
We also show the projections of the individual components of each model with
grey lines, and we compare these projections against the actual surface brightness
of the galaxies (green data points).
As a term of comparison we also show, with a dashed purple line, the
extrapolation of the \Sersic{} part of the \corser{} component.
The projections of the 1D residuals (plotted in the lower panels underneath the
surface brightness profiles) are presented both in terms of the surface brightness
difference ($\Delta\mu$), and in terms of standard deviations.
Notice that the 1D projections are centered on the galaxy center, therefore the
additional objects that we fit concurrently with A2261-BCG appear offset in this
1D representation.
For SDSS-H5, despite the fit being extended to practically the whole image
(top-right panel of Figure~\ref{figure:mosaic_and_ellipse}), the 1D projection
is limited by the largest isophote identified by IRAF.\emph{ellipse}.

\begin{figure*}
 \makebox[\linewidth]{
  \begin{overpic}[width=0.45\textwidth]
   {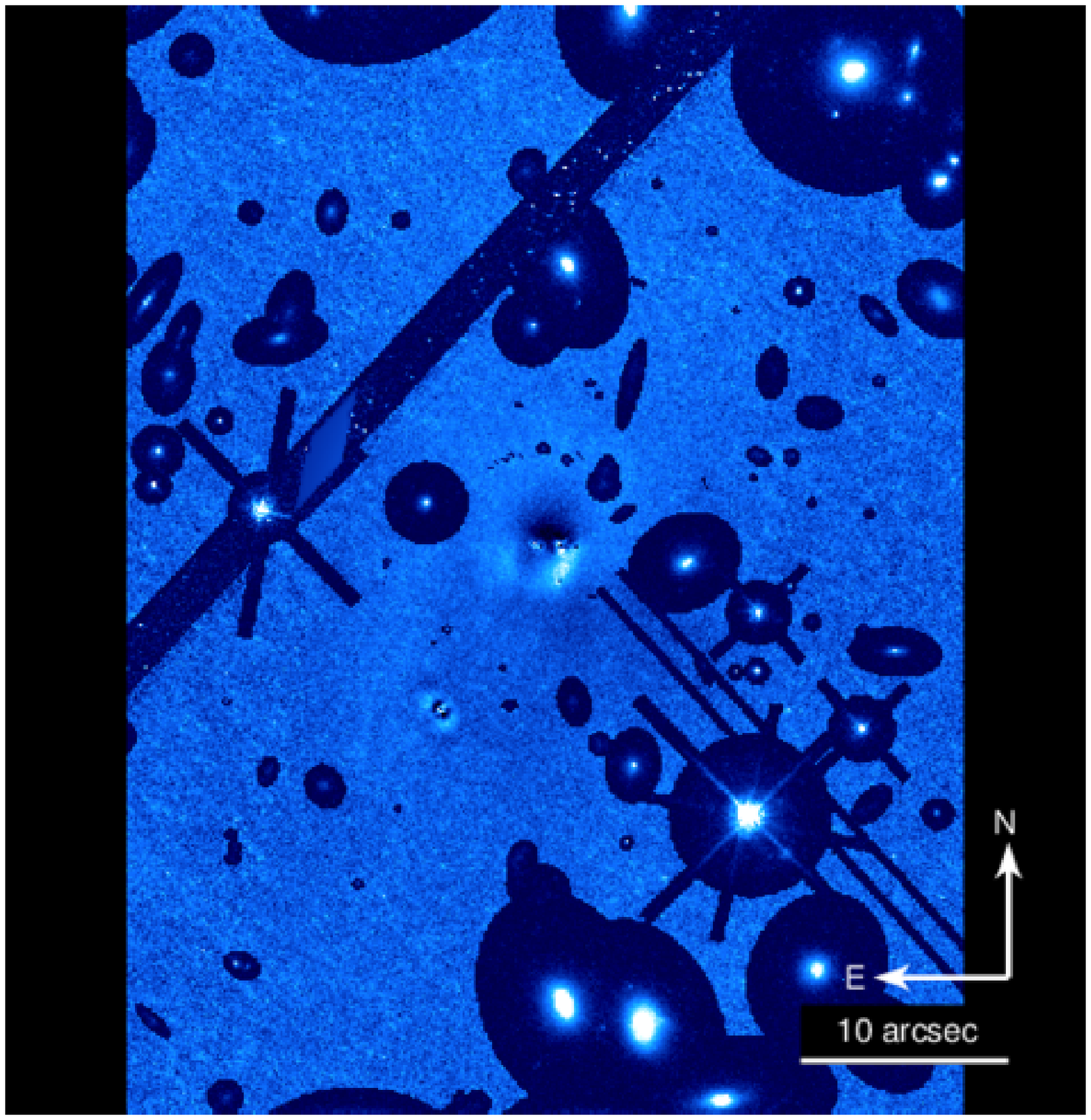}
   \put(5,5){\includegraphics[scale=0.212]{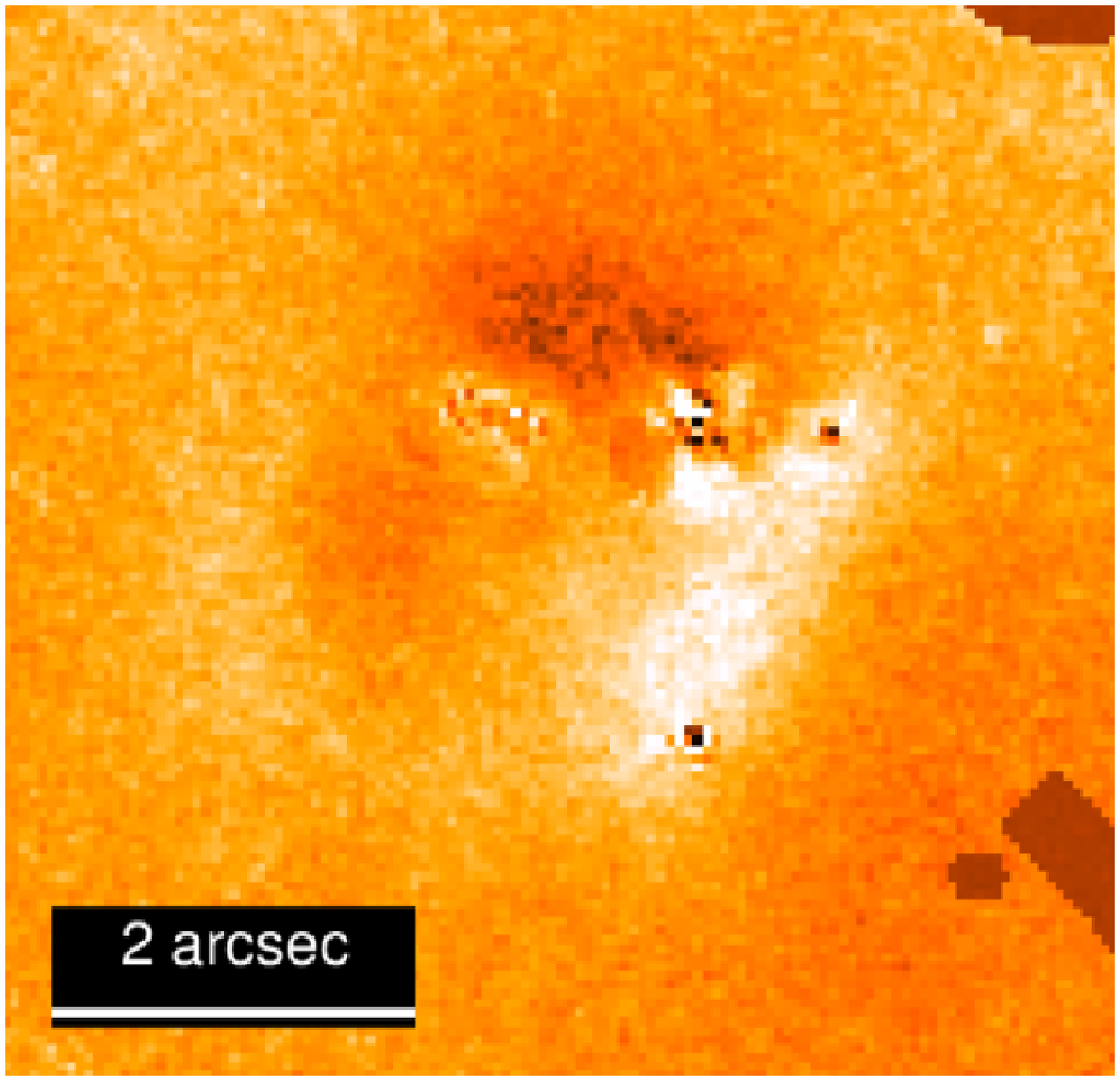}}
   \put(5,90){\fcolorbox{black}{black}{\textcolor{white}{A2261-BCG}}}
   \put(12,40){\fcolorbox{black}{black}{\textcolor{white}{A2261-BCG core}}}
  \end{overpic}

  \begin{overpic}[width=0.45\textwidth]
   {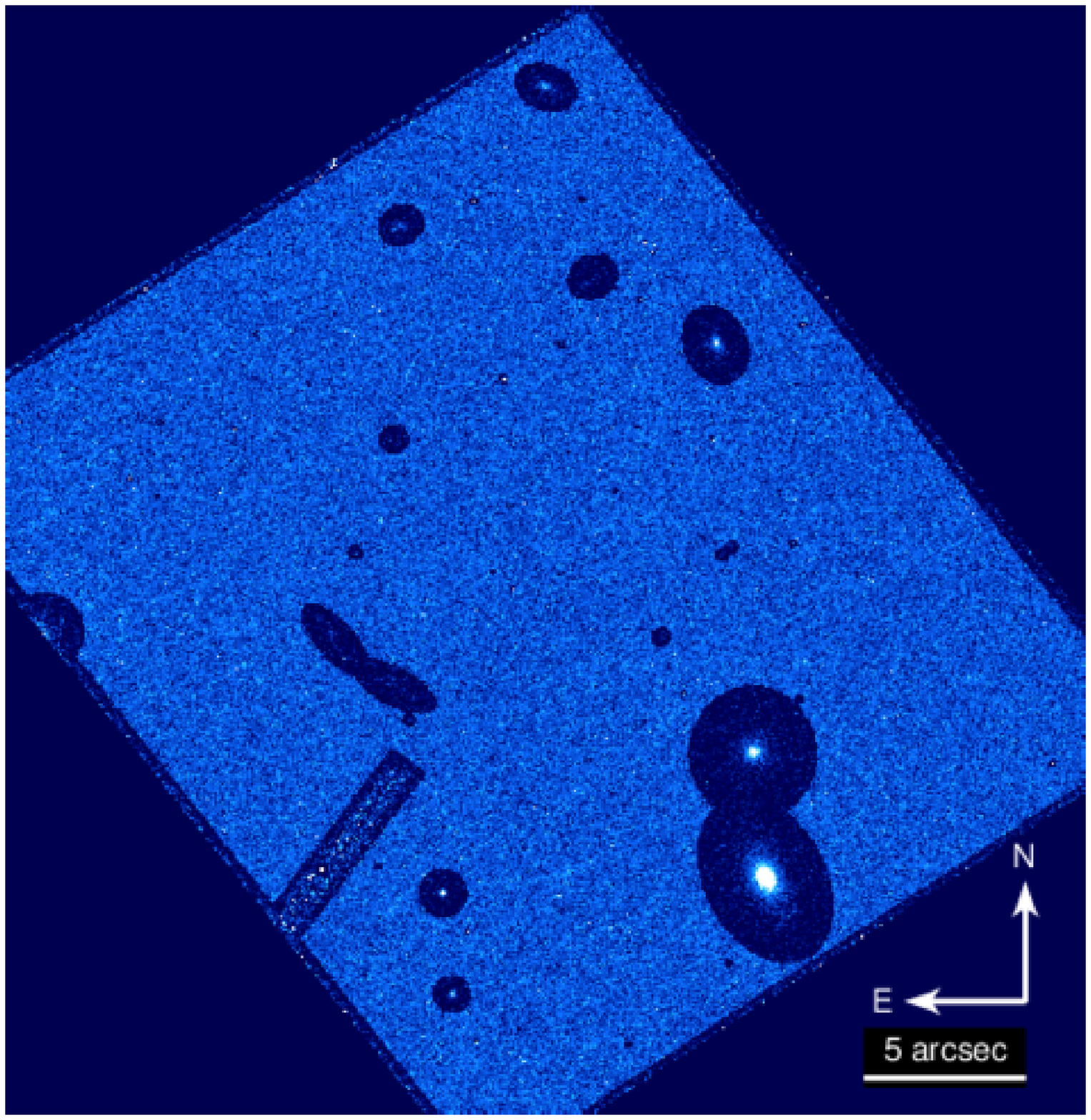}
   \put(5,90){\fcolorbox{black}{black}{\textcolor{white}{SDSS-H5}}}
  \end{overpic}
 }
 \caption{
  Residuals of the 2D fit to the image of A2261-BCG (\emph{left}) and SDSS-H5
  (\emph{right}).
  \newline
  Residual images created by first subtracting the best-fit model produced by
  \CORSAIR{}, and then normalizing by the ``sigma'' image.
  In this context, these images represent the residuals in units of standard
  deviation at each pixel position (and they are the equivalent of what we present
  in the  bottom panels of Figure \ref{figure:projection} for the 1D case).
  Masked objects have been down-scaled as in the top panels of
  Figure~\ref{figure:mosaic_and_ellipse}.
  Note: the left panel is a slightly zoomed area of that shown in the top-left
  panel of Figure~\ref{figure:mosaic_and_ellipse}.
  The insert in the left panel shows the separate fit performed on the innermost
  5$\arcsec$ square region of A2261-BCG to better constrain the models for the
  knots 1--5 (Table~\ref{table:knots}).
  The primary (dark / bright) dipole structure seen there is due to the isophote
  centroid shift, as a function of radius, seen in the lower-left panel of
  Figure~\ref{figure:mosaic_and_ellipse}.
  Note  that the contrast used in the residual images has been maximised to
  better reveal the residual patterns.
  The 2D fit (data minus model) has $\Delta_{\mu} < 0.04$~\magsb{} at all
  points, see also the 1D residual profile (Figure \ref{figure:projection}).
  \label{figure:residuals}
 }
\end{figure*}

\begin{figure*}
 \makebox[\linewidth]{
  \begin{overpic}[width=0.48\textwidth]
   {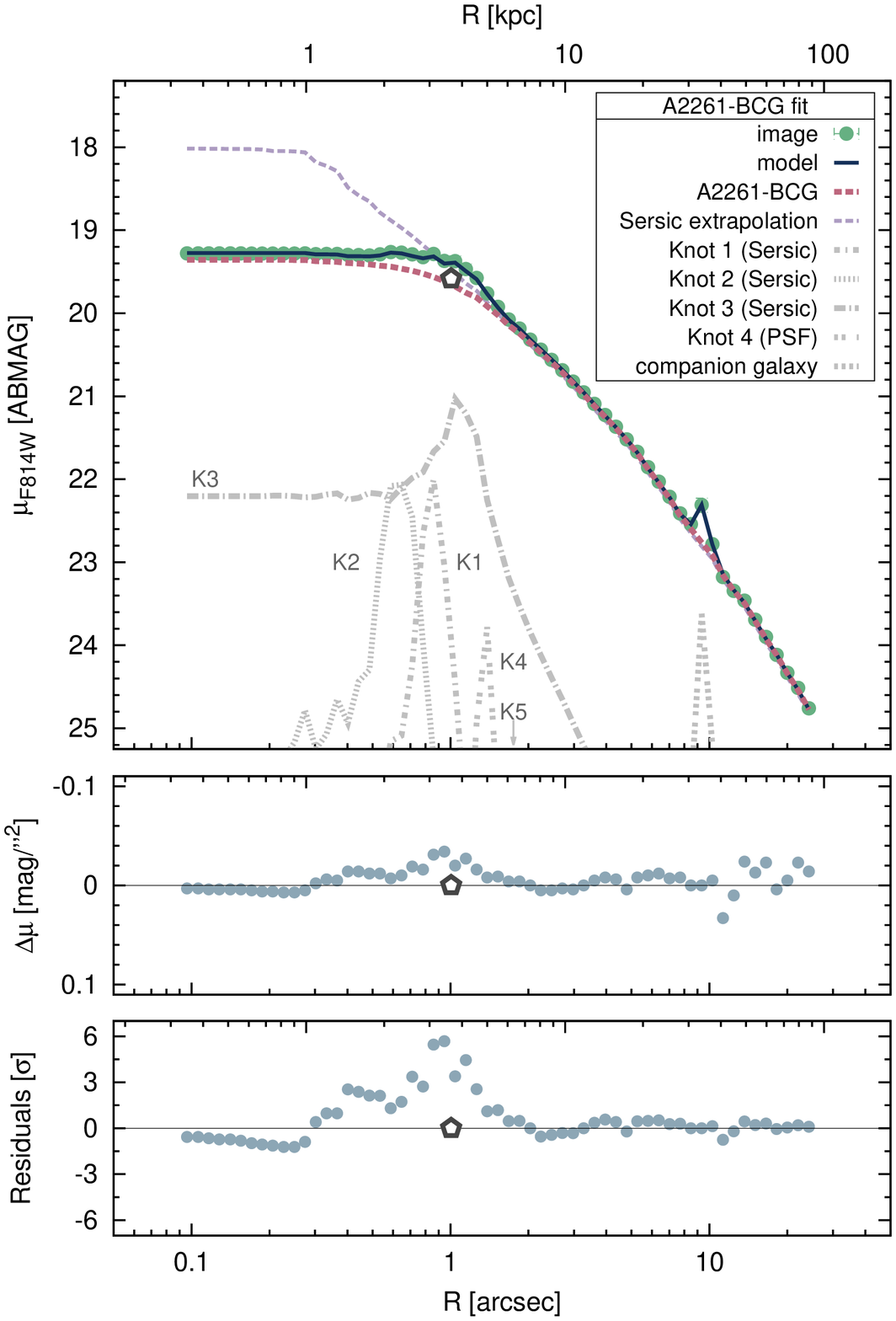}
  \end{overpic}

  \begin{overpic}[width=0.48\textwidth]
   {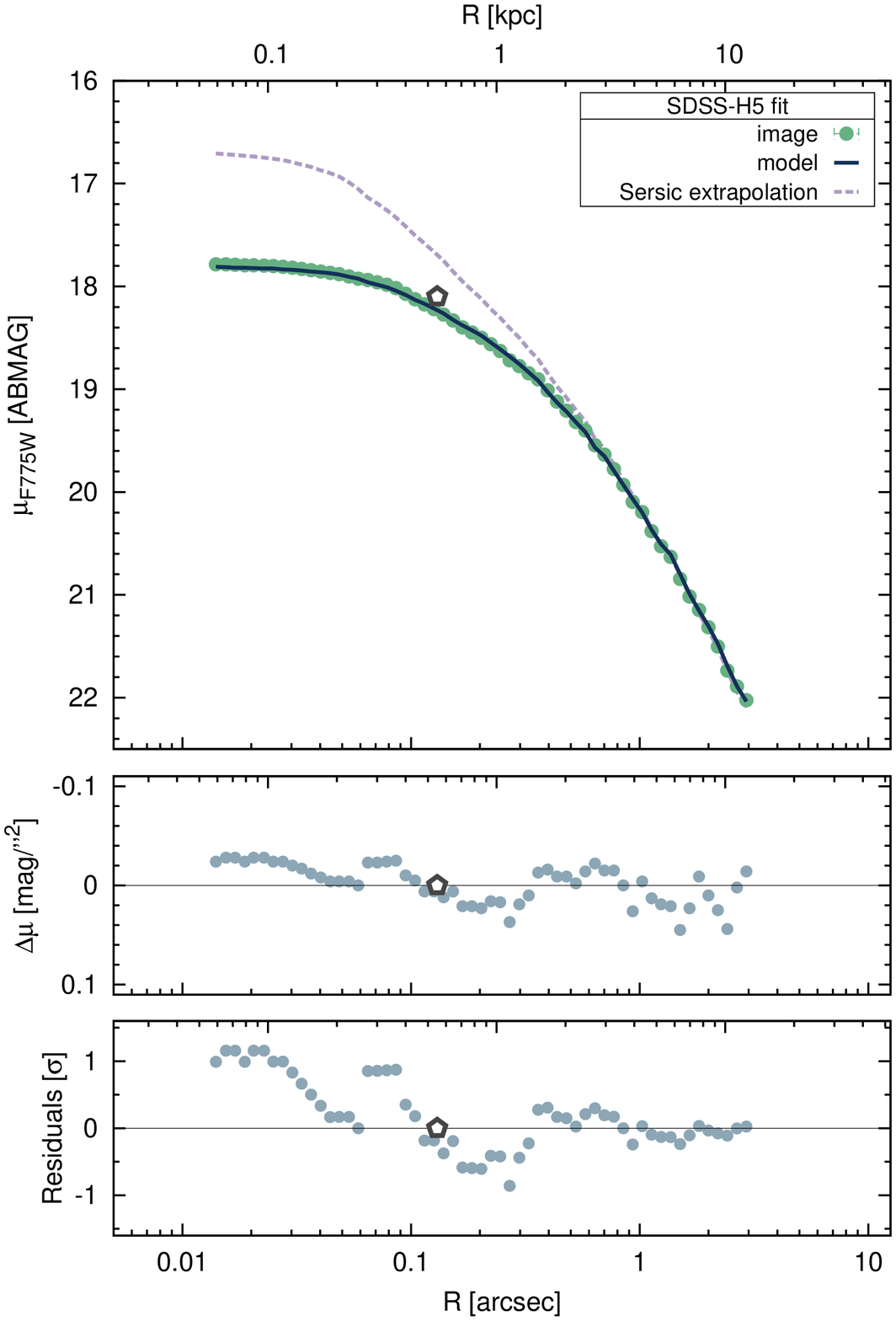}
  \end{overpic}
 }
 \caption{   
  Projections of the results of the 2D fit to the image of A2261-BCG (\emph{left})
  and SDSS-H5 (\emph{right}).
  \newline
  The green data points represent the major-axis surface brightness
  profile measured over the isophotes defined using IRAF.\emph{ellipse} (i.e.\ the
  same measurement presented in the bottom panel of
  Figure~\ref{figure:mosaic_and_ellipse}).
  The curves represent the surface brightness profile of the 2D model images measured
  over exactly the same isophotes.
  The continuous curves show the global [PSF-convolved] models, while the dashed
  curves below them represent their sub-components.
  The purple dashed lines show instead the projection of the 2D models created by
  extrapolating the \Sersic{} part of the \corser{} components.
  We stress that  all these curves are \emph{not} fits to the 1D profile, but rather
  surface brightness measurements (projections) of the 2D models, centered on the
  IRAF.\emph{ellipse} centroids.
  This is also the reason for which the profiles of some components appear as
  offset, or present ``wiggles''.  
  The projection of the surface brightness profiles of the knots 1--4 about the center
  of A2261-BCG are marked as K1--4 (the position of Knot 5 is indicated by an arrow).
  The pentagon indicates the location of the \corser{} model's break radius.
  Notice that this break radius is defined as the mid-point of the transition
  between the inner power-law and the outer \Sersic{} part of the \corser{}
  model (see the discussion in \S\ref{Fitting}).
  The panels underneath the profiles represent the data residuals about the fitted
  models, first expressed in terms of the difference in surface brightness, and
  then in terms of residuals (in units of counts) divided by the standard deviation
  as measured on the ``sigma'' image.
  The projection for SDSS-H5 is limited by the largest isophote identified by
  IRAF.\emph{ellipse}, and represents therefore only the inner  $\sim$3$\arcsec$ of
  the galaxy, while the actual 2D fit was extended to practically the whole image
  (see Figure \ref{figure:residuals}).
\label{figure:projection}
 }
\end{figure*}

We observed that, by limiting the spatial extent of the fit of A2261-BCG to the
central region, the center of the \corser{} component shifted noticeably, as expected
from the previously identified centroid shift (see \S\ref{Preliminary photometry}).
The shift of the centroid of our 2D model is obviously a function of the
fit extent; by experimenting with increasingly smaller spatial limits,
we found a shift of $\sim$0$\arcsec$.05 -- 0$\arcsec$.10, corresponding to a
physical distance of a few hundreds parsecs.

From the \corser{} fit to A2261-BCG we obtained a large break radius
$R_{b,cS}$ = 3.63~kpc, while for SDSS-H5 we obtained a more ``modest''
$R_{b,cS}$ = 0.55~kpc.
In \S\ref{Comparing our fit results with the literature} we compare these core
sizes with the results obtained in the studies of \citet[A2261-BCG]{postman} and
\citet[SDSS-H5]{hyde}, and we relate them to the
typical values observed in early-type galaxies.

\subsection[Mass deficits]{Mass deficits}
\label{Mass deficits}

The mass deficit ($M_{def}$; i.e.\ the stellar mass removed to create the core),
can be calculated from the luminosity deficit ($L_{def}$), which is defined as the
difference between the assumed pre-depletion light profile and the one actually
observed.
In the \corser{} framework, $L_{def}$ can be calculated as the difference between
the whole \corser{} model (the observed light profile), and the inward extrapolation
of its \Sersic{} part (the presumed pristine light profile).
In these regards, we remark that our measurement of the central mass deficit is
relative to the extrapolation of the outer profile.
That is, we measure the difference between the mass derived from the actual
light distribution, and what this mass would be if the spheroid's outer
\Sersic{} profile continued into the galaxy centre.
In terms of either a scoured mass or inhibited mass growth (which prevented
completion of the \Sersic{} profile over the inner radii), this approach
provides a central deficit which is equivalent to assuming that the original
light profile was described by the outer \Sersic{} model over its entire radial
extent.

As seen in \citet{graham:corser}, intermediate-luminosity galaxies such as NGC~5831
have pure S\'ersic light profiles over their entire radial extent, while fainter
ETGs often display additional nuclear components relative to their \Sersic{}
profile \citep[e.g.][]{graham:2003}. 
The luminosity deficit can then be converted into missing stellar mass assuming an
$M/L$ ratio.
To do this, we first convert the absolute magnitudes into solar luminosities,
using $M_{\odot,F814W}$ = 4.57~mag\footnote{
  \url{www.ucolick.org/~cnaw/sun.html}
}
and $M_{\odot,F775W}$ = 4.53~mag\footnote{
  \url{www.baryons.org/ezgal/filters.php}
}
 for the
absolute [AB] magnitude of the Sun.
We report the $M_{def}$ for our sample galaxies in Table~\ref{table:corser},
where we adopted the same $M/L$ used to determine the galaxy total stellar
mass (see also \S\ref{Fitting}).

The percentage of stellar mass which is displaced from the core of a 
\corser{} galaxy is typically small ($\la$~1\%) compared to the spheroid's total stellar mass,
and it therefore usually does not affect the outer stellar distribution.
In the case of A2261-BCG and SDSS-H5, the mass deficit accounts for a more
significant fraction ($\gtrsim$~4\%; see Table~\ref{table:corser}) of the total
mass of the spheroid.
Therefore in this case there could be a small additional uncertainty in using the
extrapolation of the \Sersic{} part of the \corser{} model as a proxy for the
pre-depleted light profile.
The mass deficit is high for SDSS-H5 due to the broad transition region
(i.e.\ low $\alpha$) coupled with the large \Sersic{} index.
The mass deficit relative to the predicted central SMBH mass is also
given in Table~\ref{table:corser}.
The predicted SMBH mass was obtained using the spheroid stellar mass and the
scaling relation given by \citet{scott:2013} for early-type galaxies.

\section[Discussion]{Discussion}
\label{Discussion}

\noindent
After excluding Holm~15A \citep[][]{Holm15A}, the next two largest depleted cores
reported in the literature were --- to our knowledge --- those of A2261-BCG
\citep[][]{postman}, and SDSS-H5 \citep[][]{hyde}.

\subsection[Comparing our fit results with the literature]{Comparing our fit results with the literature}
\label{Comparing our fit results with the literature}

\cite{postman} fit the 1D light profile of A2261-BCG using a Nuker model
\citep{lauer:nuker}, while \cite{hyde} used, in their analysis
of SDSS-H5, both a Nuker and a \corser{} model.
Rather than using the Nuker model ``break radius'' ($R_{b,Nuk}$) to designate the
size of a depleted core, it is now common to adopt a ``cusp radius''
$R_{\gamma\prime=0.5}$, defined as the radius at which the negative logarithmic slope
of the intensity profile ($\gamma\prime$) equals 0.5 \citep[][]{carollo}.
For simplicity, hereafter we will denote $R_{\gamma\prime=0.5}$ as
$R_{\gamma\prime}$. 
The parameter $R_{\gamma\prime}$ has been found to correlate better then $R_{b,Nuk}$
does with the galaxy properties
\citep[e.g.\ luminosity and stellar velocity dispersion; see][]{lauer:2007}.
Moreover, when there is an actual break in the profile and a partially depleted
core relative to the inward extrapolation of the outer \Sersic{} profile,
$R_{\gamma\prime}$ = $R_{b,cS}$ \citep{dullo:2013}.

\cite{postman} report a projected cusp radius\footnote{
 Throughout the paper, projected radii are expressed with a capital letter R.
}
$R_{\gamma\prime}$ of 3.2~kpc for A2261-BCG.
From their Nuker fit of SDSS-H5, \cite{hyde} reported  $R_{b,Nuk}$ = 1.574~kpc
and $R_{\gamma\prime}$ $\sim$ 0.6~kpc, while their \corser{} fit yielded a much larger
break radius $R_{b,cS}$ = 1.541~kpc, at odds with the observation that
$R_{\gamma\prime}$ usually equals $R_{b,cS}$ \citep{dullo:2014}.
According to these core sizes (Table~\ref{table:reported_cores}), A2261-BCG and
SDSS-H5 currently represent the biggest
core galaxies ever identified using a Nuker and a \corser{} model, respectively.
For a comparison, the largest cusp radii of the other known Nuker core galaxies
barely exceed $\sim$1~kpc \citep[see][their Figure 7]{postman}, while the typical
\corser{} galaxy usually hosts a core with $R_{b,cS}$ of a few hundred parsecs
\citep[e.g.][]{trujillo:corser,ferrarese:2006,richings,dullo:2012,dullo:2013,dullo:2014,rusli}.

\renewcommand{\tabcolsep}{1.0em}

\begin{deluxetable}{lccc}
 \tabletypesize{\small}
 \tablecaption{Core Sizes \label{table:reported_cores}}
 \tablehead{
  \colhead{Target}                   &
  \colhead{$R_{\gamma\prime = 0.5}$} &
  \colhead{$R_{b,cS}$}               &
  \colhead{$R_{b,cS}$}    
  \\
  \colhead{}             &
  \colhead{(literature)} &
  \colhead{(literature)} &
  \colhead{(this work)}  
  \\
  \colhead{}             &
  \colhead{[kpc]}        &
  \colhead{[kpc]}        &
  \colhead{[kpc]}                  
  \\
  \colhead{{\tiny (1)}} &
  \colhead{{\tiny (2)}} &     
  \colhead{{\tiny (3)}} &
  \colhead{{\tiny (4)}}
 }
 \startdata
A2261-BCG & 3.2 & -   & 3.63 \\
\addlinespace 
SDSS-H5   & 0.6 & 1.5 & 0.55 \\
 \enddata
 \tablecomments{
  Core sizes reported in the literature for A2261-BCG \citep{postman} and
  SDSS-H5 \citep{hyde}.
  \\
  $^{(1)}$ Target name.
  $^{(2)}$ Literature cusp radius (Nuker model).
  $^{(3)}$ Literature \corser{} break radius.
  $^{(4)}$ \corser{} break radius from Table~\ref{table:corser}.
 }
\end{deluxetable}

Our \corser{} fit to A2261-BCG confirmed that the galaxy has a large depleted
core, rather than just a shallow inner light profile. 
The luminosity deficit ($L_{def}$) that we derived as the difference between the \corser{}
profile and the inward extrapolation of its \Sersic{} part
(see \S\ref{Mass deficits}) is $\sim$ 1.43$\times$10$^{11}$ $L_{\odot}$.
This value is significantly higher than the luminosity deficit obtained by \cite{postman}
when subtracting their Nuker profile from the inward extrapolation of a \Sersic{}
model fit to their marked-by-eye ``envelope'' of A2261-BCG
($M_{V,def}$ = $-20.8$ mag, or $L_{def}$ = 1.8$\times$10$^{10}$ $L_{\odot}$).

The comparison with the results of \cite{hyde} for SDSS-H5 requires some
remarks.
The large discrepancy between their \corser{} model's $R_{b,cS}$ and their Nuker model's 
$R_{\gamma\prime}$ (see Table~\ref{table:reported_cores}), and between the $R_{e}$
values they obtained from their \corser{} model and their \Sersic{} model fit of the
galaxy (210.988~kpc and 13.907~kpc, respectively), raised our suspicion 
because when there {\it is} a depleted core, the (Nuker model)-derived value
of $R_{\gamma\prime}$ and the core-S\'ersic model's $R_{b,cS}$
are usually comparable \citep[within a factor of 2; e.g.][]{dullo:2014}, plus the
\corser{} model's effective half-light radius is basically equal to the
effective half light radius of the
extrapolation of its \Sersic{} part \citep[e.g.][]{trujillo:corser}.
Moreover, the radial extent of their fit was limited to $\sim$7$\arcsec$
(or $\sim$30~kpc; their Figure A3), corresponding to only $\sim$15\% of the
effective radius that they reported with the \corser{} model (211~kpc).
This is in part what motivated us to perform our new \corser{} fit, which 
improved on the \corser{} fit by \cite{hyde} in several aspects.

First, \cite{hyde} adopted a simplified version of the \corser{} model,
which assumed a sharp transition (i.e.\ $\alpha>>1$; their Equation 2), while
with \CORSAIR{} we were able to fit an arbitrarily smooth transition.
Given that we recover $\alpha$ $\sim$ 1.2, the assumption of a sharp transition
is not appropriate for this galaxy, and led \cite{hyde} to an excessively
large $R_{b,cS}$.  
Notice that, from the physical point of view, low-alpha \corser{} galaxies are
compatible with core excavation scenarios in the following terms.
The scouring action, whether due to an in-falling perturber (see
\S\ref{The stalled perturber scenario}) or to a binary SMBH
pair (see \S\ref{The SMBH scouring scenario}), need not occur only within the
"break radius".
The actual "loss cone" is more complicated than this, and works in the
(position,velocity) phase space.
Stars with large outer radii that plunge into the core of a galaxy can also be
cleared/scoured away, and this would broaden the transition radius, and thus a
low value of alpha need not indicate that the original profile was not described
by a \Sersic{} model.

Second, we extended the radial fit range to $\sim$18$\arcsec$
(or $\sim$75~kpc, which corresponds to $\sim$6 times our \corser{} $R_e$).
\cite{hyde} fit the surface brightness profile extracted along the major axis
of the galaxy using a PSF-deconvolved image.
The degeneracy related to the deconvolution process forced them to conservatively
exclude the innermost 0.$\arcsec$035 data, while with our 2D approach we could
virtually fit the galaxy light up to the pixel scale limit (0.$\arcsec$025;
see Table~\ref{table:sample}).
Finally, \cite{hyde} decided to apply equal weighting to each logarithmically-spaced
data point to allow the inner data points to influence their fit more. 
In our 2D fit, we were able to weight the pixel independently, following the
intrinsic Poissonian scatter (see \S\ref{PSF and sigma image}).
With this procedure we obtain a \corser{} break radius $R_{b,cS}$ = 0.6~kpc,
significantly smaller than that of \cite{hyde} (1.5~kpc), and in close agreement
with their Nuker cusp radius $R_{\gamma\prime}$ = 0.6~kpc.

\subsection[Predicting the cores size]{Predicting the cores size}
\label{Predicting the cores size}

To put our galaxies in context with respect to the population of \corser{} objects,
we can use the \mbox{$R_{b,cS}$--luminosity} relation for the spheroidal
components of \corser{} galaxies to predict the core size expected for a spheroid
with a given luminosity.
We adopted the relation by \citet[][their Table 3]{dullo:2014}:

\begin{eqnarray}
 \log(R_{b,cS}\ {pc} ) & = & (-0.45\pm0.05)(M_{V} + 22.0) +\nonumber \\
                       &   & + (1.79\pm0.06)
 \label{equation:Rb_MV}
\end{eqnarray}

\noindent
where M$_{V}$ is the absolute luminosity in the $V$ band.
For A2261-BCG, we converted our absolute, rest frame magnitude M$_{F814W,0}$ to the
$V$-band magnitude (M$_{\text{V,0}}^{\text{A2261-BCG}}$, Vega system)
using the calibrations of \citet[][their Equation 1 and Table 1]{postman},
obtaining M$_{\text{V,0}}^{\text{A2261-BCG}}$ = $-24.84$~mag.
For SDSS-H5, we estimated the rest frame $V$-band magnitude
(M$_{\text{V,0}}^{\text{SDSS-H5}}$, Vega system) by first converting our absolute,
rest frame magnitude  M$_{F775W,0}$ to the $F606W$ filter using the
calibrations of
\citet[their Table 29]{sirianni},
and then using the $F606W$--$V$ color conversions of
\citet[][their Table 3]{fukugita},
obtaining M$_{\text{V,0}}^{\text{SDSS-H5}}$ = $-23.66$~mag.

Using these magnitudes in Equation \ref{equation:Rb_MV}, we calculated an expected
core radius of log($R_{b,cS}$ [pc]) = 3.07 $_{\text{$-$0.35}}^{\text{+0.34}}$
(1.2$_{\text{$-$0.9}}^{\text{+0.9}}$~kpc) and log($R_{b,cS}$ [pc]) = 2.54
$_{\text{$-$0.33}}^{\text{+0.32}}$ (0.3$_{\text{$-$0.3}}^{\text{+0.3}}$~kpc),
for A2261-BCG and SDSS-H5 respectively\footnote{
 The uncertainty calculated from Equation \ref{equation:Rb_MV} was summed in   
 quadrature
 with the intrinsic log$(R_{b,cS})$ scatter of the data from which the equation 
 was derived
 \citep[0.3~dex; see][their Table 1]{dullo:2014}.
 Section~3.3 in \cite{graham:MBH} shows how this uncertainty is accounted for in
 the total error budget.
 The upper/lower errors on $M_{V}$ (as well as on $M_{sph}$ and $R_{b}$, used in
 \S\ref{Formation scenarios for the core}) were estimated using the simulation
 presented in the Appendix, and are reported in Table \ref{table:corser}). 
}.
Therefore, while the core of SDSS-H5 ($R_{b,cS}$ = 0.55~kpc) is compatible
(within 1-$\sigma$) with what would be expected given its spheroid luminosity, the core of
A2261-BCG ($R_{b,cS}$ = 3.63~kpc) appears somewhat larger.

\subsection[Formation scenarios for the core]{Formation scenarios for the core}
\label{Formation scenarios for the core}

\noindent
We observe that, for both galaxies, the mass deficits account for $\gtrsim$4\%
of the stellar mass of their spheroids (see Table~\ref{table:corser}), almost one
order of magnitude more than that found for the median depleted core
\citep[e.g][]{rusli,dullo:2014}.
While our results relocate SDSS-H5 into the population of \corser{} objects
with ``standard-sized'' cores ($\la$ 1 kpc), they confirm the extraordinary
large nature of the core in A2261-BCG.
In this Section, we will explore two plausible formation scenarios for the
creation of such an unusual core.

\subsubsection[The SMBH scouring scenario]{The SMBH scouring scenario}
\label{The SMBH scouring scenario}

\noindent
Several studies have highlighted a connection between the stellar mass (or light)
deficit and the mass of the central black hole \MBH{}, which suggests the
existence of an underlying link between the formation of the core and the
central SMBH
\citep[e.g.][]{begelman,faber,ravindranath,graham:2004,ferrarese:2006, lauer:2007,kormendy:2009,dullo:2014}.
These observational results have been supported by numerical simulations
in the context of the scouring SMBH binary scenario
\citep[e.g.][]{milosavljevic:2001,merritt}.
In particular, the \MBH{}--$M_{def}$ connection draws a quasi-linear relation
in the log space, although with a large scatter.
Here we test whether A2261-BCG follows this trend. 

We predicted the expected, central black hole mass in A2261-BCG (and SDSS-H5)
using our measured spheroid  
(\emph{i}) stellar mass ($M_{\rm sph,*}$) and 
(\emph{ii}) core break radius ($R_{b,cS}$). 

(i) 
We adopt the `\corser{}' \MBH{}--\Msph{} relation from \cite{scott:2013}:

\begin{eqnarray}
 \log(M_{\bullet}/M_{\odot}) & = & (9.27\pm0.09)  +\nonumber \\ 
                             &   & \hskip-40pt + (0.97\pm0.14)\log\left[ M_{\rm sph,*}/3\times 10^{11} M_{\odot} \right]. \nonumber 
 \label{equation:MBH_M}
\end{eqnarray}

\noindent
where we used the spheroid masses reported in Table~\ref{table:corser}, i.e.\
4.44$\times$10$^{12}$~\Msun{} and 1.80$\times$10$^{12}$~\Msun{} for A2261-BCG
and SDSS-H5 respectively, and we assumed an intrinsic scatter of
0.3~dex in the $\log(M_{\bullet})$ direction.
The predicted black hole masses are provided in Table~\ref{table:BH}. 

(ii)
The \MBH{}--$R_b$ relation from \citet[their Equation 13: see
also equations~15--17 in \citealp{dullo:2014}]{rusli} is such that:

\begin{eqnarray}
 \log(M_{\bullet}/M_{\odot}) & = & (10.07\pm0.16)  +\nonumber \\
                            &   & \hskip-40pt + (0.92\pm0.20)\log\left[R_b [{\rm kpc}] \right], \nonumber
 \label{equation:MBH_Rb}
\end{eqnarray}

\noindent
and is reported to have an intrinsic scatter of 0.28~dex in the
$\log(M_{\bullet})$ direction.
Assuming a 20\% uncertainty on our break radii, the predicted black hole masses
are given in Table~\ref{table:BH}.
We note that the SMBH mass predicted here for A2261-BCG will be too high if
its unusually large core was not formed via the same mechanism that formed the
cores in the spheroids that were used to construct the \MBH{}--$R_b$ relation. 

\renewcommand{\tabcolsep}{1.0em}

\begin{deluxetable}{llc}
 \tabletypesize{\small}
 \tablecaption{Predicted SMBH masses. \label{table:BH}}
 \tablehead{
  \colhead{Galaxy}          &
  \colhead{Parameter:value} &    
  \colhead{\MBH{} prediction}
  \\
  \colhead{}             &
  \colhead{} &
  \colhead{10$^9$ \Msun{}}
  \\
  \colhead{{\tiny (1)}} &
  \colhead{{\tiny (2)}} &     
  \colhead{{\tiny (3)}}
 }
 \startdata
A2261-BCG & $M_{\rm sph,*}$:  $4.44 \times 10^{12} M_{\odot}$   & $25.4^{+21.4}_{-20.7}$\\    
A2261-BCG & $r_b$: 3.63 kpc                                     & $38.5^{+30.1}_{-28.6}$\\
\addlinespace 
SDSS-H5   & $M_{\rm sph,*}$:  $1.80\times 10^{12} M_{\odot}$    & $10.6^{+8.4}_{-8.1}$ \\    
SDSS-H5   & $r_b$: 0.55 kpc                                     & $6.8^{+8.5}_{-5.1}$   \\
 \tablecomments{
  Black hole masses estimated using the scaling relations between
  \MBH{} and the spheroid's 
  stellar mass \citep[$M_{\rm sph,*}$][]{scott:2013} 
  and core break radius \citep[$R_b$][]{rusli}.
  \\
  $^{(1)}$ Target name.
  $^{(2)}$ Parameter used for calculation.
  $^{(3)}$ Predicted SMBH mass.
 }
\end{deluxetable}

In addition to these estimates, for A2261-BCG we can consider \MBH{} from the
relation of \cite{merloni} between the SMBH activity (in radio and X-ray bands)
and the SMBH mass \citep[also considered in][]{postman}.
This yields \MBH{} $\sim$ 2.0$\times$10$^{10}$~\Msun{} \citep{hlavacek:A2261_MBH}.
Notice that all of the \MBH{} reported here for A2261-BCG are comparable to, or
supersede the largest black hole mass dynamically measured
\citep[NGC~4889; \MBH{} = 2.1$\times$10$^{10}$~\Msun{};][]{mcconnell}.

We can now estimate the $M_{def}/$\MBH{} ratio (see Table~\ref{table:corser}) 
and compare this figure with the predictions of the numerical simulations
by \cite{merritt} for the SMBH scouring scenario. \cite{merritt} suggested that
$M_{def}$ $\propto$ 0.5$N$\MBH{}, where $N$ is the effective number of
major dry (i.e.\ gas poor) mergers which the galaxy experienced.
This picture would imply that A2261-BCG (and SDSS-H5) underwent some 14 consecutive dry major mergers (\emph{and} no wet merger able to refill the core), which is
more than expected from any evolutionary scenario of galaxies.
Moreover, the semi-analytical simulations of \cite{delucia} suggest that most
of the stellar mass of a BCG is assembled through \emph{minor} mergers at high
redshifts, and with very few major dry mergers.

Black holes can be responsible for the ejection of stars even
beyond the binary SMBH phase, after its coalescence, such as in the ``ejected SMBH''
scenario in which anisotropic emission of gravitational radiation results
in the SMBH being shot off in the opposite direction. 
An interesting corollary to this scenario is that the displaced SMBH might carry
along a ``cloak'' of stars \citep{merritt:2009}. 
\cite{postman} investigated this hypothesis for A2261-BCG, and suggested that knot 4
(see Figure~\ref{figure:mosaic_and_ellipse}) represents the most suitable candidate for
a stellar cloak around an ejected SMBH.
The projected isophotes of partially-depleted cores have been measured to be
rather round \citep{dullo:ellipticities}, and as such they are not suggestive
of highly eccentric orbits.
The lack of such orbits is expected to delay the merger event since the SMBH
binary takes longer to reach the close separation regime, when gravitational
radiation becomes strong enough to lead the binary to a rapid coalescence \citep{thorne,haehnelt,jaffe,sesana}.
The upshot is that there may be more mergers later on, i.e. more recently, whose
gravitational radiation we can detect today.

\subsubsection[The stalled perturber scenario]{The stalled perturber scenario}
\label{The stalled perturber scenario}

\noindent
We evaluate the ``stalled perturber'' scenario \cite[e.g.][]{goerdt} in
which the baryonic core of A2261-BCG has been excavated by a massive, compact
object which fell into the galaxy.
Clearly, in this case, knots 1--5 represent obvious candidate perturbers.
We will base our investigation on the following predictions from numerical
simulations.

It has been shown that once a flat density core is created, dynamical friction no
longer applies and the in-spiralling object will ``stall'' at the boundary of
the core \citep{goerdt:2006,read:2006a,inoue:2009,inoue:2011}.
The insert in the left panel of Figure~\ref{figure:mosaic_and_ellipse}
represents with a dashed region the extent of the core of A2261-BCG,
as we measured with our \corser{} fit, and centered in the center of our
\CORSAIR{} model fit to the inner 5$\arcsec$ region.
From the image we can only infer the de-projected position of the knots with
respect to the core center.
Therefore every object within and on the circle, nominally knots 1--3, is a possible
perturber.

It is expected that the core radius should roughly correspond to the
radius within which the enclosed (pre-depletion) mass equals the mass of the
perturber \citep{read:2006b,goerdt}.
We can approximate the original mass enclosed within $R_{b,cS}$ ($M_{enc}$) as:

\begin{equation}
 M_{enc} \sim M_{core} + M_{def,core}
 \label{equation:M_comparison}
\end{equation}

\noindent
where $M_{core}$ is the stellar mass still observed within the depleted core,
and $M_{def,core}$ is the mass deficit calculated only within $R_{b,cS}$.
We calculated $M_{core}$ from the deprojected luminosity density
profile reported in \citet[][their Figure 6]{postman}, which has been obtained
from an Abel inversion of their Nuker surface brightness profile.\footnote{
 In particular, we consider the simplest model of \cite{postman}, where
 the inner slope $\gamma$ of the Nuker model was held fixed at a value of 0.
}
According to their results, the central luminosity density ($\rho_{L,c}$)
is roughly constant ($\sim$0.03~$L_{\odot}$ pc$^{-3}$) up to the core radius.
The stellar mass still enclosed by the core is thus:

\begin{equation}
 M_{core} = {4 \over 3} \pi (R_{b,cS})^{3} \ \rho_{L,c} \ (M/L)
 \label{equation:M_core}
\end{equation}

\noindent
where $R_{b,cS}$ is our best-fit \corser{} core radius, and $M/L$ is the
same mass-to-light ratio assumed to calculate the galaxy mass (see \S\ref{Results}).
We note that this approximation smooths over the caveat that the Nuker model
overestimates the outer profile of the galaxy, hence it underestimates the
resulting central density.
In this way, we calculated $M_{core}$ =  2.1$\times$10$^{10}$~\Msun{}.
$M_{def,core}$ is 11.7$\times$10$^{10}$~\Msun{}; this value is smaller than the
$M_{def}$ reported in Table~\ref{table:corser} (17.5$\times$10$^{10}$~\Msun{})
due to our use of the mass deficit within the spherical volume bounded by
$R_{b,cS}$ and the smooth transition between the power-law and the \Sersic{} domains of
the \corser{} model (formally, the \Sersic{} extrapolation ``detaches'' from
the \corser{} profile at infinite radii).
Ultimately, these quantities yielded $M_{enc}$ = 13.8$\times$10$^{10}$~\Msun{}.

We compare this value against the mass of knot 3, the most luminous object around the
core of A2261-BCG.
Its mass ($M_{knot\ 3}$ = 4.5$\times$10$^{10}$~\Msun{}; see Table~\ref{table:knots})
is a factor of $\sim$3 smaller than $M_{enc}$.
However, the actual mass comparison should be performed against the mass
of the in-falling perturber \emph{before} the dynamical interaction, during
which the perturber itself could have lost stellar mass.
Moreover, the extended wings of knot 3 might be confused with the underlying
light of the galaxy; this may have reduced the object luminosity recovered by our fit.
Given these considerations, knot 3 represents a plausible candidate perturber
to explain the unusually large radial extent of the core in A2261-BCG. 
Knots 1 and 2 might have contributed to the excavation of the core
(although to a lesser extent) through an analogous infall process.

Finally, if the core has been excavated by an object spiralling into the
galaxy center, shredding the central cuspy distribution and producing asymmetrical
features \citep[e.g., see Figure 5 in][]{goerdt}, then the brightness centroid is
expected to shift inside the core radius, as we in fact observe
(shift $\sim$ 0$\arcsec$.05 -- 0$\arcsec$.1; see \S\ref{Fitting}).
However, a centroid displacement would be expected as well in the ``ejected SMBH''
scenario explored by \cite{postman}.

Both our results and those of \cite{postman} rule out the simple binary SMBH
scenario for the formation of the core of A2261-BCG, unless the binary has already
coalesced and the core has been formed/enlarged by the ejection of the SMBH.
As already pointed out by \cite{postman}, a core lacking its central
SMBH would be ``exposed'' to refilling from dense in-falling satellites,
implying that --- in this scenario --- the core of A2261-BCG is particularly young.

\section[Conclusions]{Conclusions}
\label{Conclusions}

\noindent
We have investigated two galaxies reported to have the largest depleted
cores according to a Nuker and a \corser{} fit, namely the BCG of
Abell~2261 \citep[A2261-BCG; $R_{\gamma\prime}$ = 3.2~kpc;][]{postman} and
SDSS-J091944.2+562201.1 \citep[SDSS-H5; $R_{b,cS}$ = 1.54 kpc;][]{hyde}, respectively.
We have modelled the two-dimensional surface brightness
distribution of the galaxies with a 2D \corser{} model, using \CORSAIR{}
\citep{corsair}.

From our fit, we obtained a \corser{} break radius $R_{b,cS}$ $\sim$ 0.55 kpc for
SDSS-H5, a value 3 times smaller than the $R_{b,cS}$ reported by \cite{hyde}, and
compatible with their cusp radius $R_{\gamma\prime}$ = 0.6~kpc.
We attribute this large discrepancy to the fact that \cite{hyde} used a
simplified formulation of the \corser{} model, assuming a sharp transition between
the \Sersic{} and the power-law, while our fit revealed that the transition
is instead very smooth
(\mbox{$\alpha$ $\sim$ 1.2}; \S\ref{Comparing our fit results with the literature}).
This led \cite{hyde} to derive an exceptionally large $R_{b,cS}$.
Given this result, we note that it would be worth re-investigating the other
galaxies in \cite{hyde}, in particular those reported to have large cores. 

We performed a \corser{} fit to A2261-BCG, and found
$R_{b,cS}$ = 3.6~kpc (\S\ref{Results}), confirming the existence of the
unusually large depleted core first claimed by 
\cite{postman} using the cusp radius.
With this $R_{b,cS}$, A2261-BCG can now be reported to have the biggest core of any
\corser{} galaxy.
The core radius of A2261-BCG is three times the value expected from the
$R_{b,cS}$--$M_{V}$ relation of \cite{dullo:2014} for \corser{} galaxies
(\S\ref{Discussion}).
We have calculated the stellar mass deficit associated with the core of A2261-BCG from
the extrapolation its outer profile, obtaining $M_{def}$ $\sim$ 1.75$\times$10$^{11}~M_{\odot}$
(\S\ref{Mass deficits}).

Moreover, we compared $M_{def}$ against the expected mass of the central black
hole (\MBH{}), which we estimated using the spheroid stellar mass 
(\S\ref{The SMBH scouring scenario}, Table~\ref{table:BH}).
The predicted black hole mass ($25\times10^9~M_{\odot}$) exceeds the most massive
SMBH mass that has ever been directly measured, and yields $M_{def}/$\MBH{} $\approx$ 7. 
This result is in disagreement with the simple (i.e.\ no gravitational-wave recoil)
SMBH binary scouring scenario given the unrealistic number of major mergers it would
imply, unless the core was subsequently enlarged by the rebound following the
ejection of the coalesced binary (\S\ref{The SMBH scouring scenario}).
Given this high $M_{def}/$\MBH{} ratio {\it and} the unusually large core radius, 
we therefore explored the ``stalled perturber'' model
\citep[e.g.][]{goerdt:2006,read:2006a,read:2006b,inoue:2009,goerdt,inoue:2011}.
According to this scheme, a captured object spiralling inward excavated the core
of A2261-BCG via dynamical friction, and finally settled at the $R_{b,cS}$ radial
distance from the galaxy center once the friction process lost efficiency.
There are several objects about the core of A2216-BCG, three of which
fall --- along the line of sight --- within the core
(``knots 1--3''; Figure \S\ref{figure:mosaic_and_ellipse}).
By testing the prediction that the mass within the core should match the
mass of the perturber \citep[e.g.][]{read:2006b,goerdt}, we suggest that
knot 3 (or a combination of the scouring actions of knots 1--3) might be
responsible for the creation of the core of A2261-BCG.

Discerning the definitive truth about A2261-BCG could benefit from a direct
(dynamical) detection, location, and measurement of the mass of its black hole,
and accurate spectroscopy of the knots about the core in order to determine their
relative velocity and age. 
In any case, our results on A2261-BCG (together with those of \citealt{postman})
pose a critical challenge to the simple binary SMBH scenario as the only
mechanism of core-formation, and calls for additional,
careful studies of high-mass core-S\'ersic galaxies. 
The ``stalled perturber'' model proved to be plausible for the explanation of the
large depleted core in A2261-BCG, and is worth consideration in future
investigations.

\acknowledgments
The authors wish to thank Tobias Goerdt for providing useful clarifications
regarding the ``stalled perturber'' scenario.
We are also very grateful to David Merritt for reviewing parts of this paper.
Finally, we thank T. Bitsakis and A. Zezas for their assistance during the review
process.
This research was supported under the Australian Research Council’s
funding scheme (FT110100263). 
This research has made use of the NASA/IPAC Extragalactic Database (NED).
Based on observations made with the NASA/ESA Hubble Space Telescope, obtained
from the data archive at the Space Telescope Science Institute. STScI is operated
by the Association of Universities for Research in Astronomy, Inc. under NASA
contract NAS 5-26555.
The HST observations are associated with GO proposal 12066 (PI: Postman, M.),
and SNAP proposal 10199 (PI: Bernardi, M.).




\appendix
\label{APPENDIX:A}

\begin{figure*}

 \definecolor{light-gray}{gray}{0.45}
 \linethickness{0.5pt}

 \makebox[\linewidth]{
  \includegraphics[]{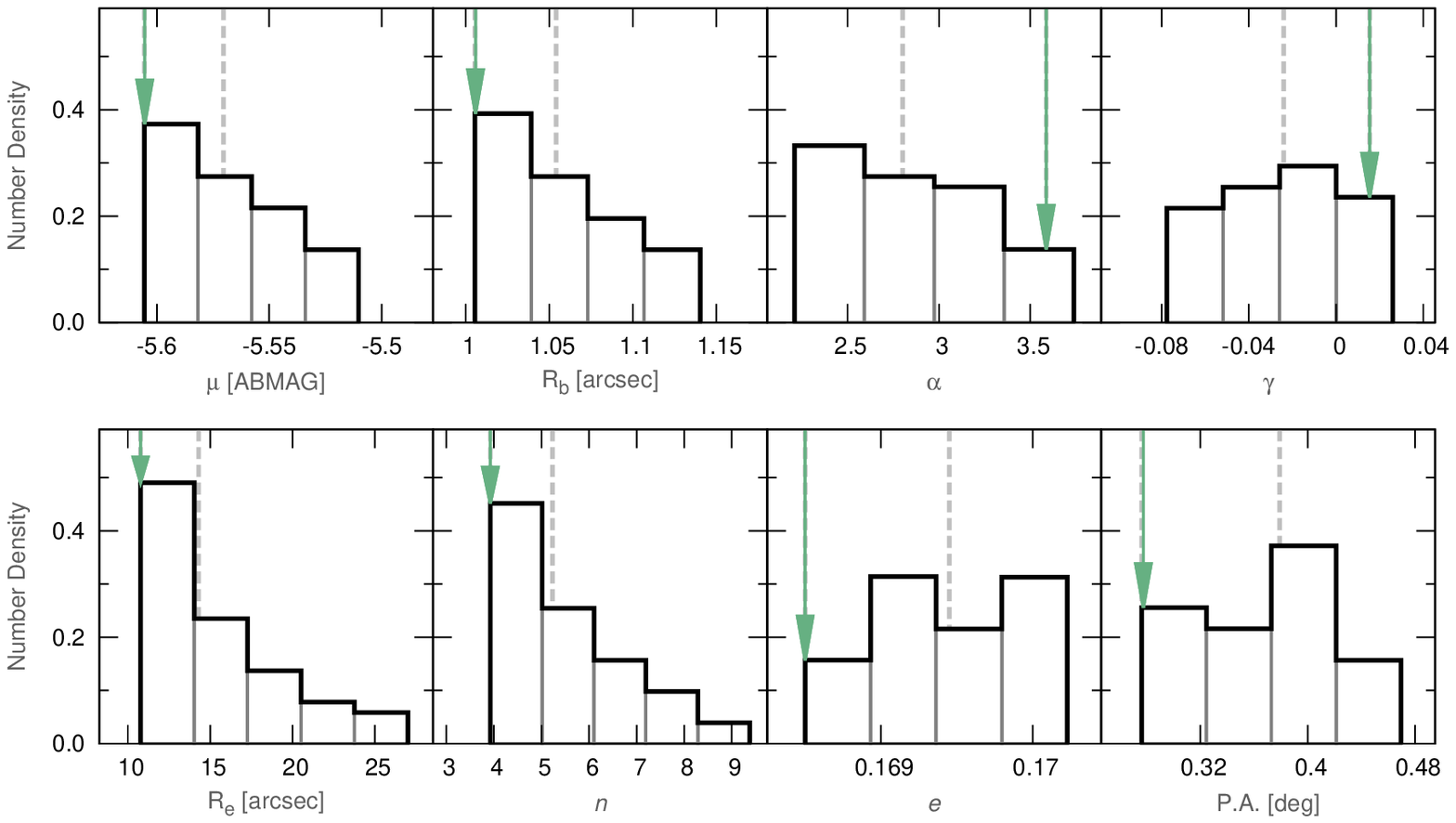}
 }
 \makebox[\linewidth]{
  \includegraphics[]{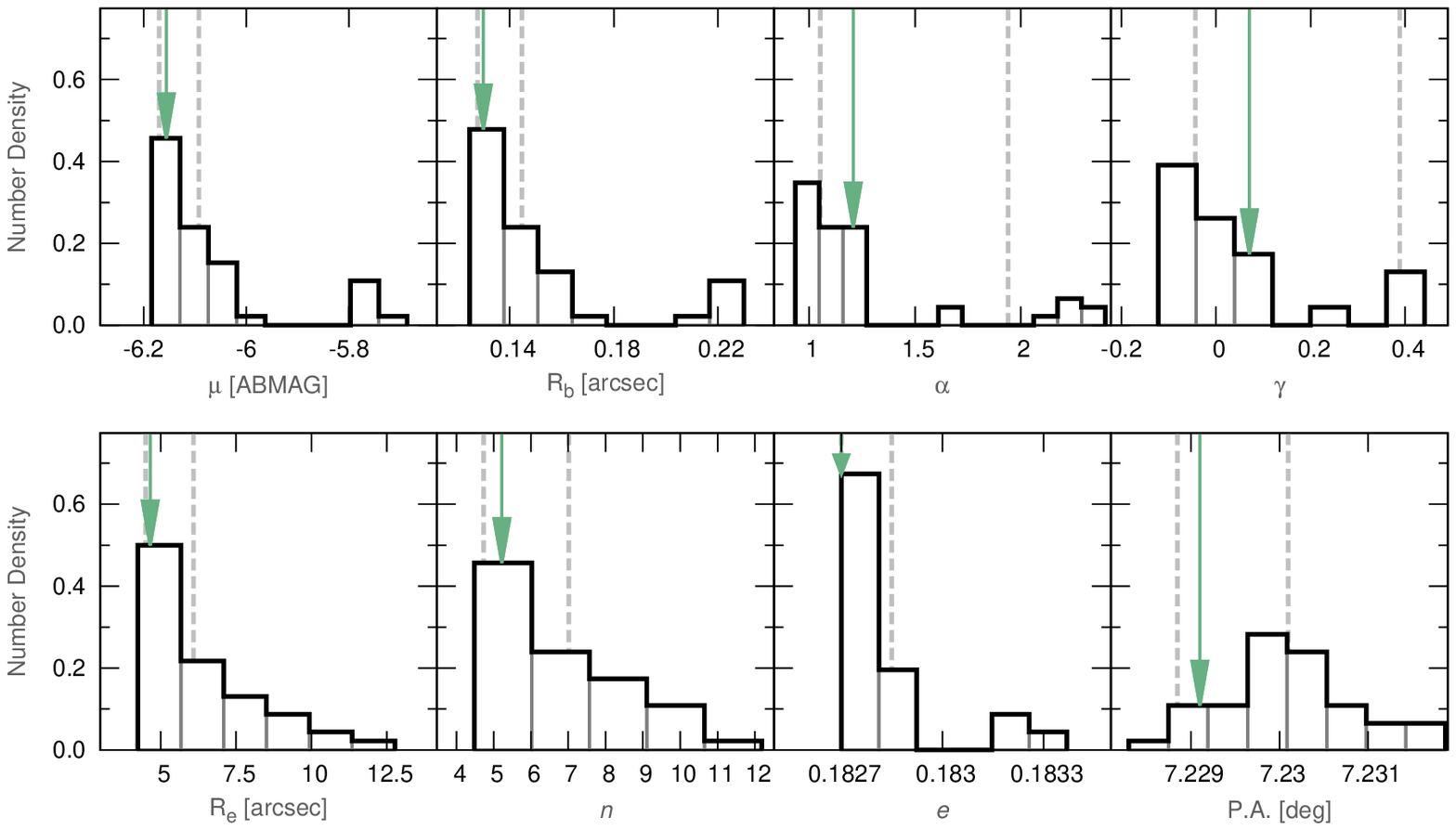}
 }
 \caption{
  Confidence Levels (CLs) for the best-fit \corser{} parameters of A2261-BCG
  (\emph{top}) and SDSS-H5 (\emph{bottom}).
  \newline
  The histograms show the probability distribution functions of the \corser{}
  parameters derived from fits performed adopting different background levels.
  The green arrows show the values of the actual best-fit parameters (i.e.,
  the ones reported in Table \ref{table:corser}), while the dashed lines represent
  the 50\% CLs around them.
  \newline
  \textsc{Note:} The fact that best-fit values (and CLs) which are upper limits
  (e.g. $\alpha$ and $\gamma$ for A2261-BCG) do not appear at the edge of the
  rightmost bin is just a representation side-effect due to the binning scheme,
  which starts from the leftmost data point, and increases with a fixed
  Freedman-Diaconis bin size.
  \label{figure:confidence_levels}
 }
\end{figure*}

\noindent
The sky background is arguably the main source of uncertainty in the 2D fit of
galaxy surface brightnesses.
In the cases in which the galaxy occupies a large fraction of the image frame,
it is virtually impossible to recover the true value of the sky, as one cannot
disentangle the contribution due to the ``wings'' of the galaxy light itself
(unless of course the intrinsic galaxy light distribution was known a-priori).
\newline
\newline
One possible approach to overcome this limitation consists in investigating the
sky values recovered from a pool of sky + galaxy fits performed at different
frame coverages
\citep[i.e. relative frame sizes with respect to the galaxy size; e.g.][]{huang}.
We preferred instead to adopt an approach which is completely independent from
the model assumed to describe the galaxy light, in order to measure the
uncertainties on the best-fit parameters due to our lack of information about
the sky.
\newline
\newline
We started from the consideration that --- in the most conservative scenario ---
the true sky value of a galaxy-dominated image could be anything between the
background value measured at the image borders (i.e. the whole background is due
to the sky brightness), and 0 (i.e. the background is totally dominated by the
galaxy light)\footnote{
 Recall that the the sky value we adopted in our best-fit (which we will label
 "assumed background" and "actual fit", respectively) was measured at the
 image borders, and hence corresponds to the assumption that the whole
 background is due to the sky.
}.
To evaluate the possible range of the fit parameters, we therefore run 50 new
fits using different fixed background values, uniformly sampled between the
aforementioned limits.
At each run, all the \corser{} parameters were left free to vary. We adopted
the same sigma image as for our actual fit in order to be able to compare the
resulting best-fit \chisq{}.
\newline
\newline
As expected, most of the fits run for background values very different from the
assumed background failed (i.e. did not converge).
For the valid fits, the distribution of each best-fit parameter was constructed
by binning the data according to the Freedman-Diaconis prescription
\citep{freedman-diaconis}, after weighting each value by the \chisq{} of the
corresponding fit.
The resulting distributions are presented in Figure
\ref{figure:confidence_levels}.
Notice that, by encompassing \emph{all} the possible sky values, these
histograms actually play the role of Probability Distribution  Functions (PDFs).
It is therefore possible to use them to calculate the Confidence Levels (CLs)
around the actual best-fit parameters (indicated by the green arrows).
The 50\% CLs are shown in the figures with dashed lines (and reported in
Table A.1).

\begin{figure*}

 \definecolor{light-gray}{gray}{0.45}
 \linethickness{0.5pt}

 \makebox[\linewidth]{
  \includegraphics[]{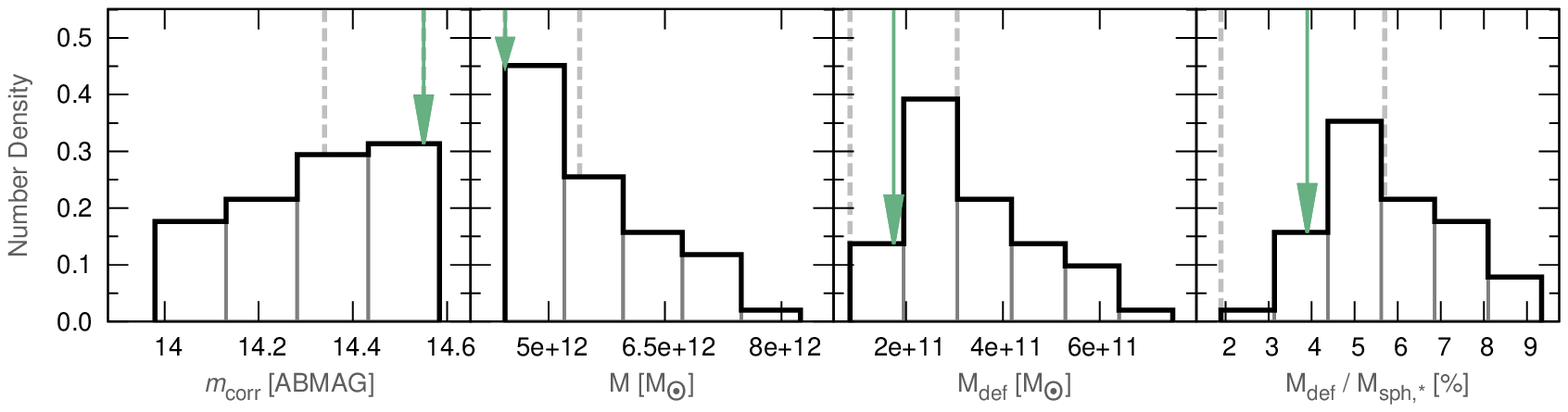}
 }
 \makebox[\linewidth]{
  \includegraphics[]{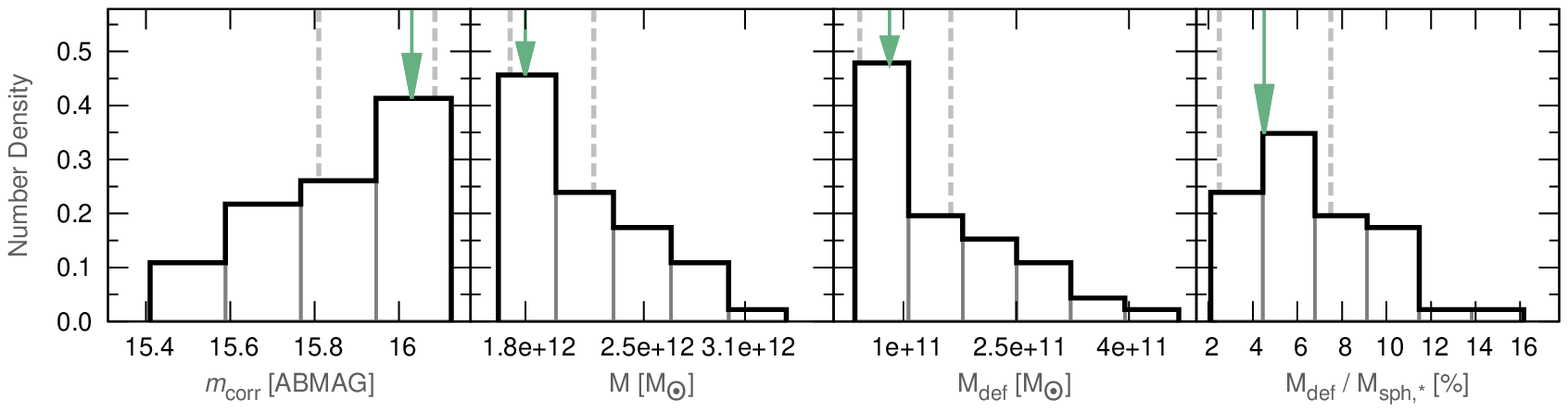}
 }
 \caption{
  Confidence Levels (CLs) for the integrated \corser{} properties of A2261-BCG
  (\emph{top}) and SDSS-H5 (\emph{bottom}).
  \newline
  The histograms represent the probability distribution functions for the (from
  left to right) extinction and K-dimming corrected apparent magnitude, spheroid
  stellar mass, mass deficit, and mass deficit to spheroid mass ratio, obtained
  simulating fits with different background levels.
  As for Figure \ref{figure:confidence_levels}, the green arrows show the values
  corresponding to our actual best-fit (Table \ref{table:corser}), while the dashed
  lines represent the 50\% CLs around them.
  \label{figure:integrated_parameters}
 }
\end{figure*}

We observe that --- as expected --- most of the parameters present a skewed PDF,
indicating a monotonic trend of the parameter with respect to the background
level.
For this reason, most of the best-fit parameters turn out to be
upper/lower limits.
In particular, the \Sersic{} index and the effective radius
appear to show the stronger dependence on the background\footnote{
 Notice that on top of the background dependence, the \Sersic{} index and the
 effective radius of the \Sersic{} model also present a true covariance 
 \citep[e.g.\ ][]{trujillo:2001}.
}.
The parameters which are obviously insensitive to the background level, such
as the P.A. and axis ratio, vary stochastically among the simulated fits, and
hence present flatter distributions with smaller relative widths.
\newline
\newline
Finally, in order to determine the uncertainties upon the integrated quantities
(total magnitude, mass, mass deficit, and mass deficit to total mass ratio) we
adopted a similar approach.
Nominally, we calculated those quantities for each of the simulated fit, and
then adopted the 50\% CLs around each [best-fit] quantity as an estimate of its
error.
The probability distributions for the integrated quantities are shown in the
Figure \ref{figure:integrated_parameters}.

\end{document}